\documentclass{elsart}

\usepackage{natbib}

\usepackage{graphicx}

\usepackage{amssymb}

\begin{document}

\begin{frontmatter}



\title{Global modelling of the early Martian climate under a denser CO$_2$ atmosphere: Water cycle and ice evolution}


\author[LMD]{R. Wordsworth\thanksref{CHI}}
\author[LMD]{F. Forget}
\author[LMD]{E. Millour}
\author[BRO]{J.~W. Head}
\author[LMD]{J.-B. Madeleine\thanksref{BRO}}
\author[LMD]{B. Charnay}
\address[LMD]{Laboratoire de M\'et\'erologie Dynamique, Institut Pierre Simon Laplace, Paris, France.}
\address[BRO]{Department of Geological Sciences, Brown University, Providence 02912 RI, USA.}
\thanks[CHI]{Current address: Department of the Geological Sciences, University of Chicago, 60637 IL, USA.}



\begin{abstract}
We discuss 3D global simulations of the early Martian climate that we have performed assuming a faint young Sun and denser CO$_2$ atmosphere. We include a self-consistent representation of the water cycle, with atmosphere-surface interactions, atmospheric transport, and the radiative effects of CO$_2$ and H$_2$O gas and clouds taken into account. We find that for atmospheric pressures greater than a fraction of a bar, the adiabatic cooling effect causes temperatures in the southern highland valley network regions to fall significantly below the global average. Long-term climate evolution simulations indicate that in these circumstances, water ice is transported to the highlands from low-lying regions for a wide range of orbital obliquities, regardless of the extent of the Tharsis bulge. In addition, an extended water ice cap forms on the southern pole, approximately corresponding to the location of the Noachian/Hesperian era Dorsa Argentea Formation. Even for a multiple-bar CO$_2$ atmosphere, conditions are too cold to allow long-term surface liquid water. Limited melting occurs on warm summer days in some locations, but only for surface albedo and thermal inertia conditions that may be unrealistic for water ice. Nonetheless, meteorite impacts and volcanism could potentially cause intense episodic melting under such conditions. Because ice migration to higher altitudes is a robust mechanism for recharging highland water sources after such events, we suggest that this globally sub-zero, `icy highlands' scenario for the late Noachian climate may be sufficient to explain most of the fluvial geology without the need to invoke additional long-term warming mechanisms or an early warm, wet Mars.
\end{abstract}


\end{frontmatter}

\section{Introduction}
\label{sec:intro}

After many decades of observational and theoretical research, the nature of the early Martian climate remains an essentially unsolved problem. Extensive geological evidence indicates that there was both flowing liquid water (e.g., \cite{Carr1995,Irwin2005,Fassett2008a,Hynek2010}) and standing bodies of water (e.g., \cite{Fassett2008a}) on the Martian surface in the late Noachian, but a comprehensive, integrated explanation for the observations remains elusive. As the young Sun was fainter by around 25~\% in the Noachian (before approx. 3.5 GYa) \citep{Gough1981}, to date no climate model has been able to produce long-term warm, wet conditions in this period convincingly. Transient warming events have been proposed to explain some of the observations, but there is still no consensus as to their rate of occurrence or overall importance.

The geomorphological evidence for an altered climate on early Mars includes extensive dendritic channels across the highland Noachian terrain (the famous `valley networks') \citep{Carr:96,Fassett2008a,Hynek2010}, fossilized river deltas with meandering features \citep{Mali:03,Fassett2005}, records of quasi-periodic sediment deposition \citep{Lewis2008}, and regions of enhanced erosion most readily explained through fluvial activity \citep{Hynek2001}. Some studies have also suggested evidence for an ancient ocean in the low-lying northern plains. These include a global analysis of the martian hydrosphere \citep{Clifford2001} and an assessment of river delta / valley network contact altitudes \citep{diAchille2010}. However, in the absence of other evidence, the existence of a northern ocean in the Noachian remains highly controversial.

More recent geochemical evidence of aqueous alteration on Mars has both broadened and complicated our view of the early climate. Observations by the OMEGA and CRISM instruments on the Mars Express / Mars Reconnaissance Orbiter spacecraft \citep{Poulet2005,Bibring2006,Mustard2008,Ehlmann2011} showed widespread evidence for phyllosilicate ($\sim$clay) and sulphate minerals across the central and southern Noachian terrain. Surface aqueous minerals are rarer in Mars' northern lowlands, which are mostly covered by younger Hesperian-era lava plains \citep{Head2002,Salvatore2010} and outflow channel effluent \citep{Kreslavsky2002}. However, phyllosilicates have been detected in some large northern impact craters that penetrated through these later deposits \citep{Carter2010}. As these impacts were understood to have excavated ancient Noachian terrain from below the lava plains, it seems plausible that aqueous alteration was once widespread in both hemispheres, on or just beneath the Martian surface.

Evidence from crater counting \citep{Fassett2008,Fassett2011} suggests that valley network formation was active during the Noachian but ended near the Noachian-Hesperian boundary (approx. 3.5 GYa in their analysis). Broadly speaking, this period overlaps with the end of the period when impacts were frequent and the Tharsis rise was still forming. Interestingly, however, crater statistics also suggest that the main period of phyllosilicate formation ended somewhat before the last valley networks were created. Few Late Noachian open-basin lakes \citep{Fassett2008a} show evidence of extensive in-situ phyllosilicates on their floors \citep{Goudge2012} and in those that do, the clays appear to have been transported there from older deposits by way of valley networks \citep[e.g., ][]{Ehlmann2008}.

Interpreting the surface conditions necessary to form the observed phyllosilicates on Mars remains a key challenge to understanding the Noachian climate. It is clear that there is substantial diversity in the early Martian mineralogical record, which probably at least partially reflects progressive changes in environmental conditions over time \citep{Bibring2006,Mustard2008,Murchie2009,AndrewsHanna2010,AndrewsHanna2011}. Nonetheless, the most recent reviews of the available geochemical evidence suggest that the majority of phyllosilicate deposits may have been formed via subsurface hydrothermal alteration \citep{Ehlmann2011} or episodic processes, as opposed to a long-term, warm wet climate.

While it is likely that Mars once possessed a thicker CO$_2$ atmosphere than it has today, it is well known that CO$_2$ gaseous absorption alone cannot produce a greenhouse effect strong enough to allow liquid water on early Mars at any atmospheric pressure \citep{Kasting1991}. Various alternative explanations for an early warm, wet climate have been put forward. Two of the most notable are additional absorption by volcanically emitted sulphur dioxide \citep{Halevy2007}, and downward scattering of outgoing infrared radiation by CO$_2$ clouds \citep{Forget1997}. However, both these hypotheses have been criticised as insufficient in later studies \citep{Colaprete2003,Tian2010}. Sulphur-induced warming is attractive due to the correlation between the timing of Tharsis formation and the valley networks and the abundance of sulphate minerals on the Martian surface. However, \cite{Tian2010} argued that this mechanism would be ineffective on timescales longer than a few months due to the cooling effects of sulphate aerosol formation in the high atmosphere. CO$_2$ clouds are a robust feature of cold CO$_2$ atmospheres that have already been observed in the mesosphere of present-day Mars \citep{Montmessin2007}. They can cause extremely effective warming via infrared scattering if they form at an optimal altitude and have global coverage close to 100~\%. However, our 3D simulations of dry CO$_2$ atmospheres \citep{Forget2012} suggest that their warming effect is unlikely to be strong enough to raise global mean temperatures above the melting point of water for reasonable atmospheric pressures.

Given the problems with steady-state warm, wet models, other researchers have proposed that extreme events such as meteorite impacts could be capable of causing enough warming to explain the observed erosion alone \citep{Segura2002,Segura2008,Toon2010}. These authors proposed that transient steam atmospheres would form for up to several millenia as a result of impacts between 30 and 250 km in diameter. They argued that the enhanced precipitation rates under such conditions would be sufficient to carve valley networks similar to those observed on Mars, and hence that a long-term warm climate was not necessary to explain the geological evidence. This hypothesis has been questioned by later studies -- for example, landform evolution modelling of the Parana Valles region (-20$^\circ$ N, 15$^\circ$ W) \citep{Barnhart2009} has suggested that the near-absence of crater rim breaches there is indicative of a long-term, semi-arid climate, as opposed to intermittent catastrophic deluges. Other researchers have argued that with realistic values of soil erodability, there is a significant discrepancy (of order 10$^4$) between the estimated Noachian erosion rates and the total erosion possible due to post-impact rainfall (Jim Kasting, private communication). Hence impact-generated steam atmospheres alone still appear unable to explain key elements of the geological observations, and the role of impacts in the Noachian hydrological cycle in general remains unclear.

Most previous theoretical studies of the early Martian climate have used one-dimensional, globally averaged models. While such models have the advantage of allowing a simple and rapid assessment of warming for a given atmosphere, they are incapable of addressing the influence of seasonal and topographic temperature variations on the global water cycle. \cite{StewartJ2008} examined the impact of sulphur volatiles on climate in a 3D general circulation model (GCM), but they did not include a dynamic water cycle or the radiative effects of clouds or aerosols. To our knowledge, no other study has yet attempted to model the primitive Martian climate in a 3D GCM.

Here we describe a range of three-dimensional simulations we have performed to investigate possible climate scenarios on early Mars. Our approach has been to study only the simplest possible atmospheric compositions, but to treat all physical processes as accurately as possible. We modelled the early Martian climate in 3D under a denser CO$_2$ atmosphere with a) dynamical representation of cloud formation and radiative effects (CO$_2$ and H$_2$O) b) self-consistent, integrated representation of the water cycle and c) accurate parameterisation of dense CO$_2$  radiative transfer. We have studied the effects of varying atmospheric pressure, orbital obliquity, surface topography and starting H$_2$O inventory. In a companion paper \citep{Forget2012}, we describe the climate under dry (pure CO$_2$) conditions. Here we focus on the water cycle, including its effects on global climate and long-term surface ice stabilization. Based on our results, we propose a new hypothesis for valley network formation that combines aspects of previous steady-state and transient warming theories.

In Section \ref{sec:method}, we describe our climate model, including the radiative transfer and dynamical modules and assumptions on the water cycle and cloud formation. In Section \ref{sec:results}, we describe the results. First, 100\% relative humidity simulations are analysed and compared with results assuming a dry atmosphere \citep{Forget2012}. Next, simulations with a self-consistent water cycle and varying assumptions on the initial CO$_2$ and H$_2$O inventories and surface topography are described. Particular emphasis is placed on a) the long-term evolution of the global hydrology towards a steady state and b) local melting due to short-term transient heating events. In Section \ref{sec:discuss} we discuss our results in the context of constraints from geological observations and atmospheric evolution theory, and assess the probable effects of impacts during a period of higher flux. Finally, we describe what we view as the most likely scenario for valley network formation in the late Noachian and suggest a few directions for future study.

\section{Method}\label{sec:method}

To produce our results we used the LMD Generic Climate Model, a new climate simulator with generalised radiative transfer and cloud physics that we have developed for a range of planetary applications \citep{Wordsworth2010b,Selsis2011,Wordsworth2011a}. The model uses the LMDZ 3D dynamical core \citep{Hourdin2006}, which is based on a finite-difference formulation of the classical primitive equations of meteorology. The numerical scheme is formulated to conserve both potential enstrophy and total angular momentum \citep{Sadourny1975}.

Scale-selective hyperdiffusion was used in the horizontal plane for stability, and linear damping was applied in the topmost levels to eliminate the artificial reflection of vertically propagating waves. The planetary boundary layer was parameterised using the method of \cite{Mellor1982} and \cite{Galperin1988} to calculate turbulent mixing, with the latent heat of H$_2$O also taken into account in the surface temperature calculations when surface ice / water was present. As in \cite{Wordsworth2011a}, a standard roughness coefficient of $z_0=1\times10^{-2}$ m was used for both rocky and ice / water surfaces. A spatial resolution of $32\times32\times15$ in longitude, latitude and altitude was used for the simulations. This was slightly greater than the $32\times24\times15$ used in most simulations in \cite{Forget2012} to allow more accurate representation of the latitudinal transport of water vapour, but still low enough to allow long-term simulation of system evolution in reasonable computation times. 

\subsection{Radiative transfer}

Our radiative transfer scheme was similar to that used in previous studies \citep[e.g., ][]{Wordsworth2011a}. For a given mixture of atmospheric gases, we computed high resolution spectra over a range of temperatures, pressures and gas mixing ratios using the HITRAN 2008 database \citep{Rothman2009}. For this study we used a 6 $\times$ 9 $\times$ 7 temperature, pressure and H$_2$O volume mixing ratio grid with values $T = \{100, 150,\ldots, 350 \}$ K, $p  = \{ 10^{-3}, 10^{-2}, \ldots, 10^5 \} $ mbar and $q_{H_2O}=\{10^{-7}, 10^{-6}, \ldots, 10^{-1} \}$, respectively. The correlated-$k$ method was used to produce a smaller database of coefficients suitable for fast calculation in a GCM. The model used 32 spectral bands in the longwave and 36 in the shortwave, and sixteen points for the $g$-space integration, where $g$ is the cumulated distribution function of the absorption data for each band. CO$_2$ collision-induced absorption was included using a parameterisation based on the most recent theoretical and experimental studies \citep{Wordsworth2010,Gruszka1998,Baranov2004}. H$_2$O lines were truncated at 25~cm$^{-1}$, while the H$_2$O continuum was included using the CKD model \citep{Clough1989}. In clear-sky one-dimensional radiative-convective tests (results not shown) the H$_2$O continuum  was found to cause an increase of less than 1~K in the global mean surface temperatures under early Mars conditions. A two-stream scheme \citep{toon1989} was used to account for the radiative effects of aerosols (here, CO$_2$ and H$_2$O clouds) and Rayleigh scattering, as in \citet{Wordsworth2010b}. The hemispheric mean method was used in the infrared, while $\delta$-quadrature was used in the visible. Finally, a solar flux of 441.1 W~m$^{-2}$ was used (75$\%$ of the present-day value), corresponding to the reduced luminosity 3.8~Ga based on standard solar evolution models \citep{Gough1981}. For a discussion of the uncertainties regarding the evolution of solar luminosity with time, see \cite{Forget2012}.

\subsection{Water cycle and CO$_2$ clouds}

Three tracer species were used in the simulations: CO$_2$ ice, H$_2$O ice and H$_2$O vapour. Tracers were advected in the atmosphere, mixed by turbulence and convection, and subjected to changes due to sublimation / evaporation and condensation and interaction with the surface. For both gases, condensation was assumed to occur when the atmospheric temperature dropped below the saturation temperature. Local mean CO$_2$ and H$_2$O cloud particle sizes were determined from the amount of condensed material and the number density of cloud condensation nuclei $[CCN]$. The latter parameter was taken to be constant everywhere in the atmosphere. Its effect on cloud radiative effects and hence on climate is discussed in detail in \cite{Forget2012}; here it was taken to be a global constant for CO$_2$ clouds ($10^5$ kg$^{-1}$; see Table \ref{tab:params}).

Ice particles of both species were sedimented according to a version of Stokes' law appropriate for a wide range of pressures \citep{Rossow1978}. Below the stratosphere, adjustment was used to relax temperatures due to convection and / or condensation of CO$_2$ and H$_2$O.  For H$_2$O, moist and large-scale convection were taken into account following \cite{Manabe1967}. Precipitation of H$_2$O was parameterized using a simple cloud water content threshold $l_0$ \citep{Emanuel1999}, with precipitation evaporation also taken into account. Tests using the generic model applied to the present-day Earth showed that this scheme with the value of $l_0$ given in Table \ref{tab:params} reasonably reproduced the observed cloud radiative effects there. To test the sensitivity of our results to the H$_2$O cloud assumptions, we performed a range of simulations where we varied the assumed value of  $l_0$ and $[CCN]$ for H$_2$O clouds. The results of these tests are described in Section \ref{subsec:watercycle}.

On the surface, the local albedo was varied according to the composition (rocky, ocean, CO$_2$ or H$_2$O ice; see Table \ref{tab:params}). When the surface ice amount $q_{ice}$ was lower than a threshold value  $q_{covered} = 33$ kg m$^{-2}$
(corresponding to a uniform layer of approx. 3.5~cm thickness; \cite{LeTreut1991}), partial coverage was assumed and the albedo $A_s$ was calculated from the rock and water ice albedos $A_r$ and $A_i$ as
\begin{equation}
A_s(\theta, \lambda) = A_r + (A_i - A_r) \frac{q_{ice}}{q_{covered}} \label{eq:albedo}
\end{equation}
where $\theta$ is longitude and $\lambda$ is latitude. For $q_{ice}>q_{covered}$, the local albedo was set to $A_i$. When CO$_2$ ice formed, the local albedo was set to $A_i$ immediately, as in \cite{Forget2012}. Water ice formation (melting) was assumed to occur when the surface temperature was lower (higher) than 273~K, and the effects of the latent heat of fusion on temperature were taken into account. As in most simulations only transient melting occurred in localised regions, the effects of runoff were not taken into account. Surface temperatures were computed from radiative, sensible and latent surface heat fluxes using an 18-level model of thermal diffusion in the soil \citep{Forget1999} and assuming a homogeneous thermal inertia (250 J~m$^{-2}$~s$^{-1\slash 2}$~K$^{-1}$). The dependence of the results on the assumption of constant thermal inertia are discussed in Section \ref{sec:results}.

\subsection{Ice equilibration algorithm}\label{subsec:icealgol}

The time taken for an atmosphere to reach thermal equilibrium can be estimated from the radiative relaxation timescale \citep{GoodyYung1989}
\begin{equation}
\tau_r = \frac{c_p p_{s}}{\sigma g T_{e}^3},
\end{equation}
where $c_p$, $p_s$, g, $T_e$ and $\sigma$ are the specific heat capacity of the atmosphere, the mean surface pressure, the surface gravity, the atmospheric emission temperature and the Stefan-Boltzmann constant, respectively. Taking $T_e=200$ K, $c_p = 800$ J K$^{-1}$ kg$^{-1}$, $p_s = 1$ bar and $g=3.72$ m s$^{-2}$ yields $\tau_r \sim 1$ Mars year. For simulations without a water cycle, this was a good approximation to the equilibration time for the entire climate.

When a water cycle is present, equilibration times can be far longer. Particularly for cold climates, where sublimation and light snowfall are the dominant forms of water transport, the surface H$_2$O distribution can take thousands of years or more to reach a steady state. Running the 3D model took around 5 hours for one simulated Mars year at the chosen resolution, so evaluating climate evolution in the standard mode was prohibitively time-consuming.

To resolve this problem, we implemented an ice iteration scheme. Starting from the initial surface ice distribution, we ran the GCM normally in intervals of two years. In the second year, we evaluated the annual mean ice rate of change $\langle \partial h_{ice} \slash \partial t \rangle$ at each gridpoint. This rate was then multiplied by a multiple-year timestep $\Delta t$ to give the updated surface ice distribution
\begin{equation}
h_{ice}^+ = h_{ice} + \langle{\frac{\partial h_{ice}}{\partial t}\rangle} \Delta t. \label{eq:icealgol}
\end{equation}
In addition, all ice was removed in regions where $h_{ice}^+$ dropped below zero or ice coverage was seasonal (i.e. $h_{ice}=0$ at some point during the year).  After redistribution, the amount in each cell was normalised to conserve the total ice mass in the system $M_{H2O}$. The GCM was then run again for two years and the process repeated until a steady state was achieved.

Trial and error showed that correct choice of the timestep $\Delta t$ was important: when it was too high, the surface ice distribution tended to fluctuate and did not reach a steady state, while when it was too low the simulations took a prohibitively long time to converge. We used a variable timestep to produce the results described here. For the first 5 iterations, $\Delta t$ was set to 100 years, after which it was reduced to 10 years for the final 15 iterations. This allowed us to access the final state of the climate system after reasonable computation times without compromising the accuracy of the results.

In all simulations we set the total ice mass $M_{H2O} = 4 \pi R^2  q_{covered}$, with $R$ the planetary radius. This quantity was chosen so that for completely even ice coverage, the thickness would be 3.5~cm: just enough for the entire surface to have the maximum albedo $A_i$ [see equation (\ref{eq:albedo})]. While the total Martian H$_2$O inventory in the Noachian is likely to have been significantly greater than this, such an approach allows us to study the influence of topography and climate on the steady state of the system without using unreasonably long iteration times. It is also conservative, in the sense that a greater total H$_2$O inventory would allow more ice to accumulate in cold-trap regions and hence more potential melting due to seasonal variations or transient events (see Sections \ref{subsec:transmelt} and \ref{sec:discuss}).
Two types of initial conditions were used: in the first, ice was restricted to altitudes lower than -4~km from the datum (`icy lowlands'), while in the second, ice was restricted to latitudes of magnitude greater than 80$^\circ$ (`icy poles'). Varying the initial conditions in this way allowed us to study the uniqueness of climate equilibria reached using equation (\ref{eq:icealgol}).

\subsection{Topography}

\begin{figure}[h]
	\begin{center}
		{\includegraphics[width=2.5in]{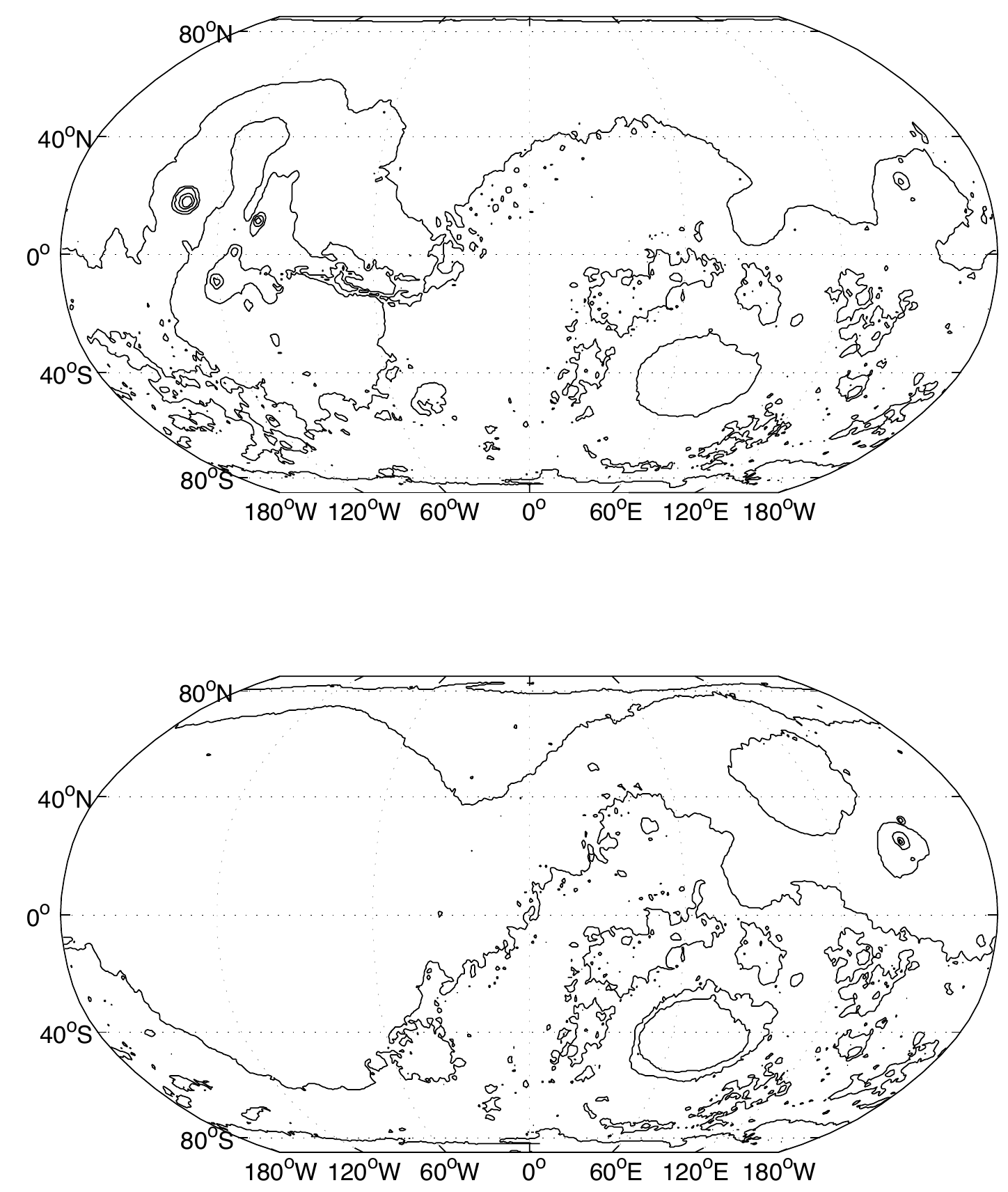}}
	\end{center}
	\caption{Comparison of present-day Martian topography (top) with the reduced Tharsis bulge topography (bottom) described by equation (\ref{eq:nuketharsis}) and used in some simulations.}
	\label{fig:topocompare}
\end{figure}

Techniques such as spherical elastic shell modelling and statistical crater analysis date the formation of the majority of the Tharsis rise to the mid to late Noachian \citep{Phillips2001,Fassett2011}. Here, we performed most simulations with present-day surface topography, but we also investigated the climatic effects of a reduced Tharsis bulge. For these cases, we used the formula
\begin{equation}
\phi_{mod}(\theta,\lambda)=\phi - (\phi + 4000 g) \mbox{exp}[ -(\chi \slash 65^\circ)^{4.5}],\label{eq:nuketharsis}
\end{equation}
to convert the present-day geopotential $\phi$ to $\phi_{mod}$, where $\chi = ([\theta+100^\circ]^2+\lambda^2)^{0.5}$ and $g$ is surface gravity. Figure \ref{fig:topocompare} compares contour plots of the standard topography and that described by equation (\ref{eq:nuketharsis}).

\begin{table}[h]
\centering
\caption{Standard parameters used in the climate simulations.}
\begin{tabular}{ll}
\hline
\hline
Parameter & Values \\
\hline
Solar flux $F_s$ [W~m$^{-2}$] & 441.1 \\
Orbital eccentricity  $e$ & 0.0 \\
Obliquity $\phi$ [degrees]  & 25.0, 45.0 \\
Surface albedo (rocky)  $A_r$ & 0.2 \\
Surface albedo (liquid water)  $A_w$ & 0.07 \\
Surface albedo (CO$_2$ / H$_2$O ice)  $A_i$ & 0.5 \\
Surface topography   & present-day, no Tharsis \\
Initial CO$_2$ partial pressure  $p_{CO_2}$ [bars] &  0.008 to 2 \\
Surface roughness coefficient  $z_0$ [m] & $1\times10^{-2}$ \\
Surf. therm. inertia  $\mathcal I$ [J~m$^{-2}$~s$^{-1\slash 2}$~K$^{-1}$] & $250$ \\
H$_2$O precipitation threshold  $l_0$ [kg kg$^{-1}$] & 0.001 \\
No. of cloud condens. nuclei  $[CCN]$ [kg$^{-1}$] & $1\times10^5$ \\
\hline \hline
\end{tabular}\label{tab:params}
\end{table}

\section{Results}\label{sec:results}

\subsection{Fixed atmospheric humidity simulations}
\label{subsec:thermalresults}

In \cite{Forget2012}, we described the climate of early Mars with a dense, dry (pure CO$_2$) atmosphere. Here we begin by considering the effects of water vapour on surface temperatures. We performed simulations as in the baseline cases of \cite{Forget2012} (variable CO$_2$ pressure, circular orbit, 25$^\circ$ obliquity, present-day topography) for a) pure CO$_2$ atmospheres and b) atmospheres with relative humidity of 1.0 at all altitudes. In these initial simulations, water vapour was only included in the atmospheric radiative transfer, and not treated as a dynamical tracer or used in convective lapse rate calculations. Neglecting water vapour in lapse rate calculations should only create a small positive surface temperature error in cold climates, and the effect of water clouds on climate in our simulations with a full water cycle was small (this is discussed further later). This approach hence allowed us to place an upper bound on the possible greenhouse warming in our calculations, independent of uncertainties in H$_2$O convection parameterisations.

\begin{figure}[h]
	\begin{center}
		{\includegraphics[width=4in]{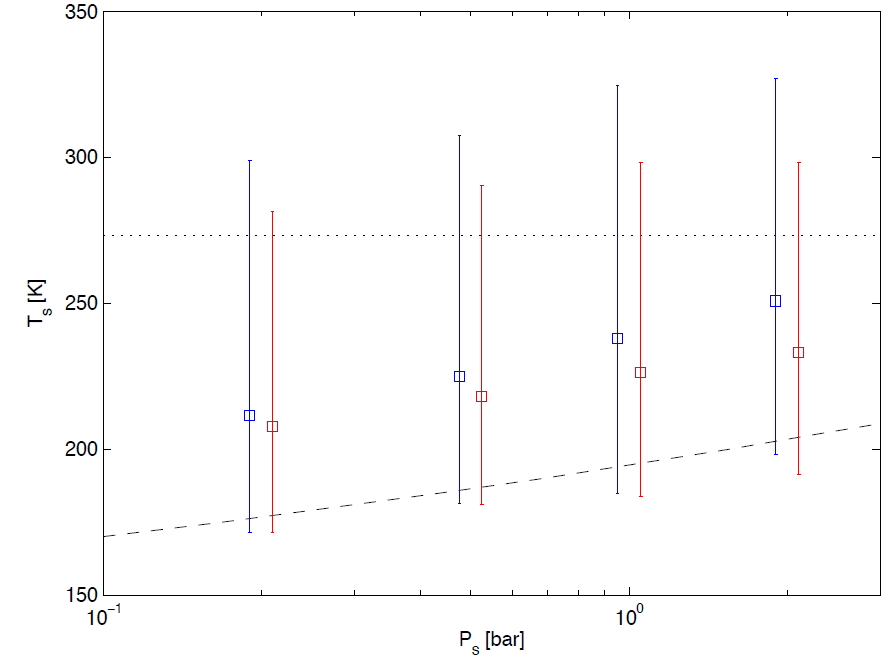}}
	\end{center}
	\caption{Effects of atmospheric CO$_2$ and H$_2$O on global temperature. Error bars show mean and maximum / minimum surface temperature vs. pressure (sampled over one orbit and across the surface) for dry CO$_2$ atmospheres (red), and simulations with 100\% relative humidity (blue) but no H$_2$O clouds. Dashed and dotted black lines show the condensation curve of CO$_2$ and the melting point of H$_2$O, respectively. For this plot simulations were performed at 0.2, 0.5, 1 and 2~bar; the dry and wet data are slightly separated for clarity only. }
\label{fig:tsurf1D}
\end{figure}

In Figure \ref{fig:tsurf1D}, global mean and annual maximum/minimum temperatures from simulations with surface pressures ranging from 0.2 to 2 bar are plotted. Comparison with results for clear CO$_2$ atmospheres shows the warming effect of CO$_2$ clouds (see Fig. 1 of \cite{Forget2012}). As expected, in atmospheres saturated with water vapour the net warming is greater (by a factor of a few Kelvin to $\sim$20~K, depending on the surface pressure). Although there are no scenarios where the mean surface temperature exceeds 273~K, annual maximum temperatures are significantly greater than this in all cases. For pressures greater than 2 bar or less than 0.5 bar, permanent CO$_2$ ice caps appeared in the simulations, indicating the onset of atmospheric collapse. As discussed in \cite{Forget2012}, finding the equilibrium state in such cases would require extremely long iteration times. Between 0.5 and 2 bar, however, CO$_2$ surface ice was seasonal, and the climate hence reached a steady state on a timescale of order $\tau_r$ (in the absence of a full water cycle).

\begin{figure}[h]
	\begin{center}
		{\includegraphics[width=4.0in]{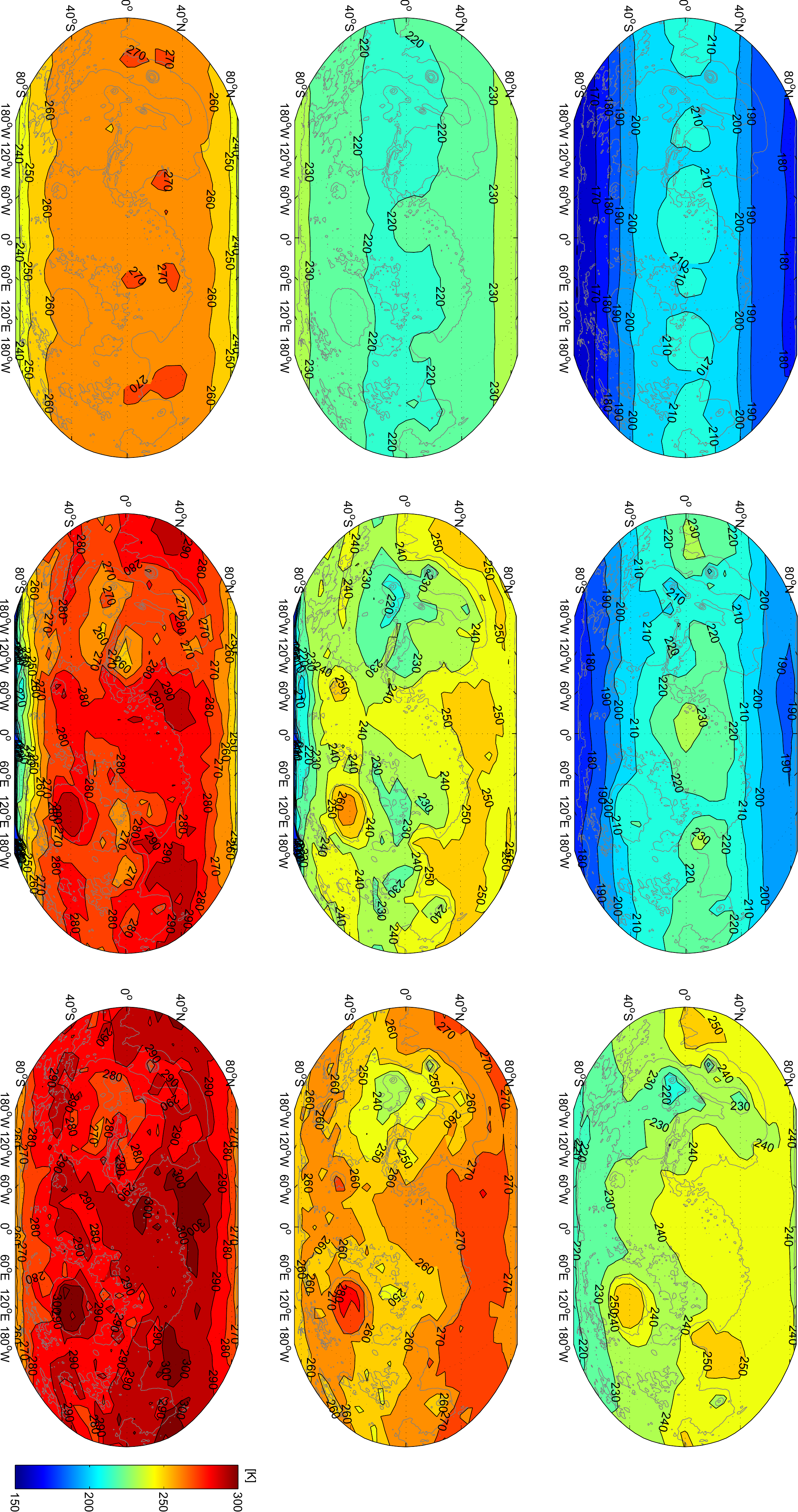}}
	\end{center}
	\caption{Surface temperature mean and variations in the 100\% relative humidity simulations. Top, middle and bottom plots show the annual mean, the diurnally averaged annual maximum, and the absolute annual maximum, respectively. Left, right and center columns are for 0.008, 0.2 and 1~bar, respectively. Black contours show the topography used in the simulation. Strong correlation of annual mean surface temperature with altitude (as occurs on e.g. present-day Earth and Venus) is apparent in the 0.2 and 1~bar cases.}
\label{fig:tsurf2D}
\end{figure}

In Figure \ref{fig:tsurf2D}, annual mean (top row), diurnally averaged annual maximum (seasonal maximum; middle row) and absolute annual maximum (bottom row) surface temperatures in 2D are plotted for atmospheres of 0.008 (left), 0.2 (middle) and 1 (right) bar pressure in the maximally H$_2$O-saturated simulations in equilibrium. The diurnal averaging consisted of a running one-day mean over results sampled 8 times per day. The adiabatic cooling effect described in \cite{Forget2012} is clearly apparent in the annual mean temperatures (Figure \ref{fig:tsurf2D}, top row): at 0.008 and 0.2~bar (left and middle) the main temperature difference is between the poles and the equator, while at 1~bar (right) altitude-temperature correlations dominate. With an atmospheric pressure close to that of Earth, the regions of Mars with the most concentrated evidence for flowing liquid water (the southern Noachian highlands) are among the coldest on the planet.

The seasonal and absolute maximum temperatures are partially correlated with altitude even at 0.2~bar (Figure \ref{fig:tsurf2D}, middle column, middle and bottom), with the result that only small regions of the southern highlands are ever warmer than 273~K at that pressure. At 1~bar, seasonal maximum temperatures (Figure \ref{fig:tsurf2D}, right column, middle) are only above 273~K in the northern plains, parts of Arabia Terra and the Hellas and Argyre basins. However, the absolute maximum temperatures exceed 273~K across most of the planet (Figure \ref{fig:tsurf2D}, right column, bottom) except the Tharsis rise and southern pole. This has interesting implications for the transient melting of water ice, as we describe in the next section.

\subsection{Simulations with a water cycle}
\label{subsec:watercycle}

When the water cycle is treated self-consistently, surface temperatures can differ from those described in the last section due to a) variations in atmospheric water vapour content, b) the radiative effect of water clouds and c) albedo changes due to surface ice and water. We studied the evolution of the surface ice distribution under these conditions using the ice evolution algorithm described in Section \ref{sec:method}. We tested the effects of obliquity, the surface topography and the starting ice distribution in the model.

In general terms, ice should tend to accumulate over time in regions where it is most stable; i.e., the coldest parts of the planet. On present-day Mars, the atmospheric influence on the surface energy balance is small, and the annual mean surface temperature is primarily a function of latitude \cite[e.g.,][]{Forget1999}. However, as discussed in the previous section, increased thermal coupling between the atmosphere and the surface at higher pressures in the past \citep{Forget2012} would have caused a significantly greater correlation between temperature and altitude, suggesting that ice might have accumulated preferentially in highland regions then. In reality, atmospheric dynamics can also play an important role on the H$_2$O surface distribution. This can occur through processes such as  topographic forcing of precipitation or the creation of global bands of convergence / divergence (c.f. the inter-tropical convergence zone on Earth, e.g. \cite{Pierrehumbert2011BOOK}). The complexity of these effects is an important reason why 3D circulation modelling is required for a self-consistent analysis of the primitive Martian water cycle.

\subsubsection{Surface ice evolution}
\begin{figure}[h]
	\begin{center}
		{\includegraphics[width=2.25in]{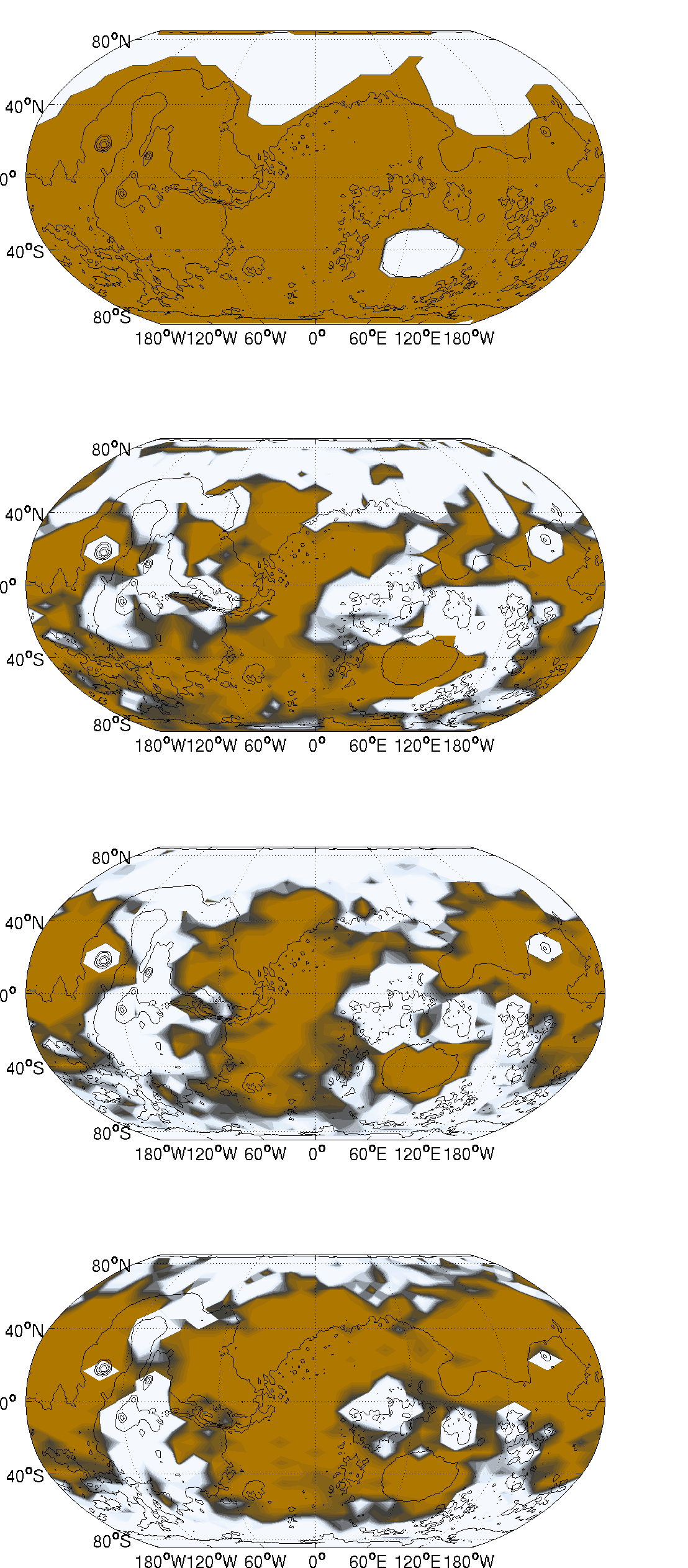}}
		{\includegraphics[width=2.25in]{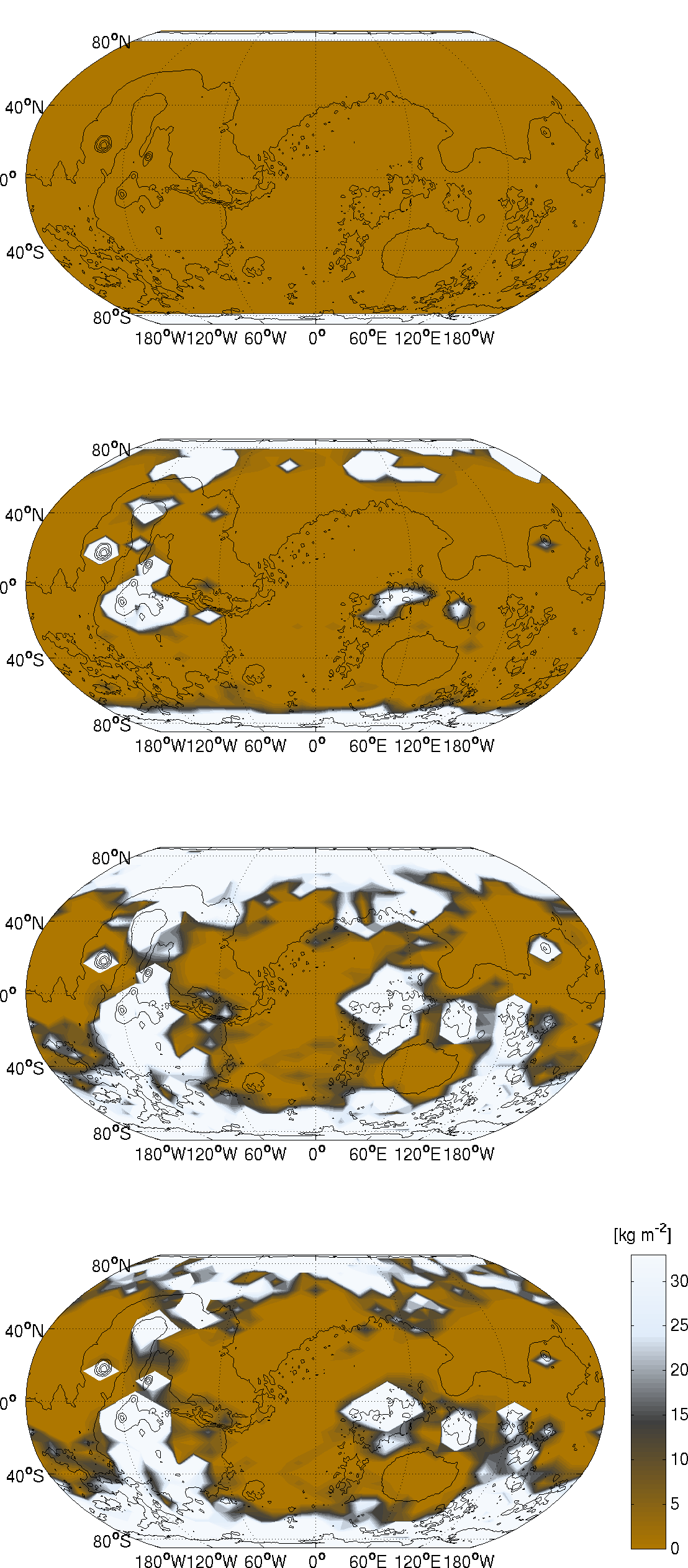}}
	\end{center}
	\caption{Evolution of H$_2$O ice in simulations using the iteration scheme described in Section \ref{sec:method}. From top to bottom, plots show surface ice density in kg m$^{-2}$ at the start of the simulation and annual mean after 4, 20 and 40 Mars years. Ice iteration was performed every two years, with a 100-year timestep used for the first 5 iterations and 10-year timesteps used thereafter. Simulations were performed at 1~bar mean surface pressure with obliquity 25$^\circ$. Left and right columns show cases with initial ice deposits in low-lying regions and at the poles, respectively.}
	\label{fig:iceevolplots1}
\end{figure}

\begin{figure}[h]
	\begin{center}
		{\includegraphics[height=2.1in]{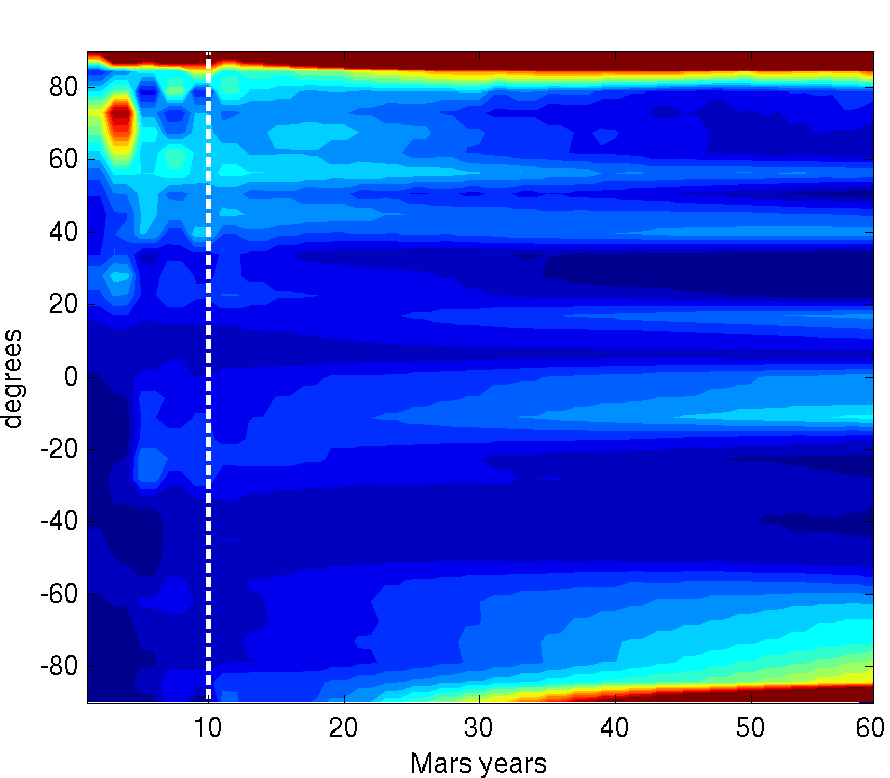}}
		{\includegraphics[height=2.1in]{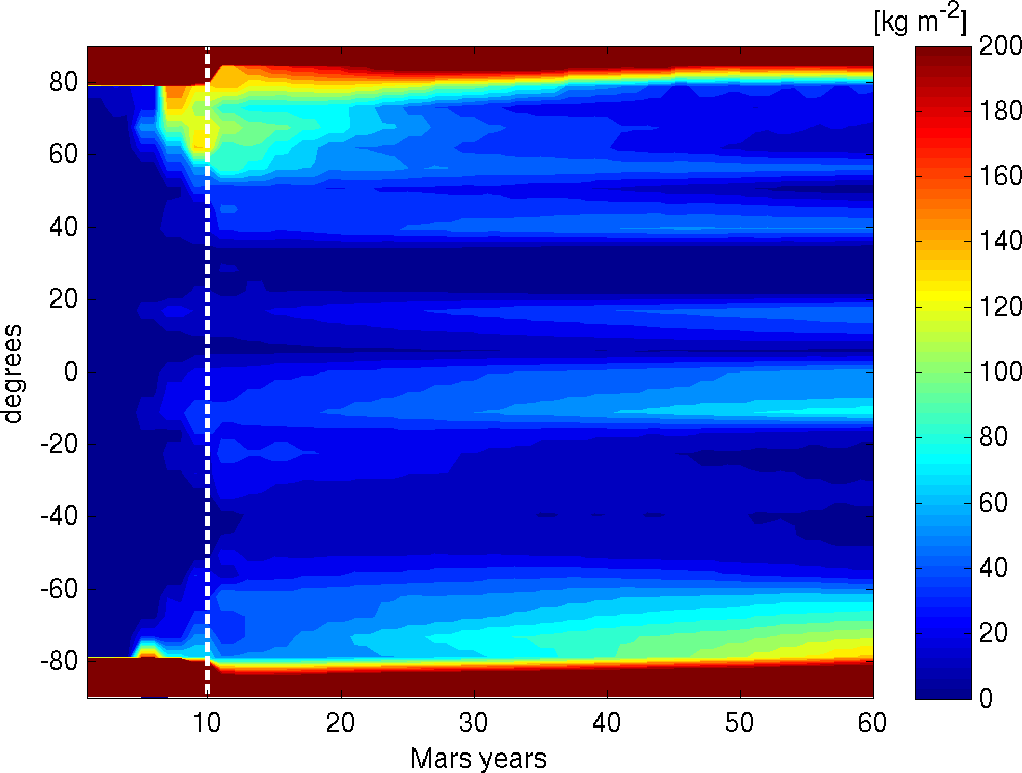}}
	\end{center}
	\caption{Zonally and yearly averaged H$_2$O surface ice density as a function of latitude and time for the same simulations as in Figure \ref{fig:iceevolplots1}. The dotted vertical line indicates the transition from 100-year to 10-year timesteps in the ice evolution algorithm.}
	\label{fig:iceevolavged}
\end{figure}

\begin{figure}[h]
	\begin{center}
		{\includegraphics[width=2.25in]{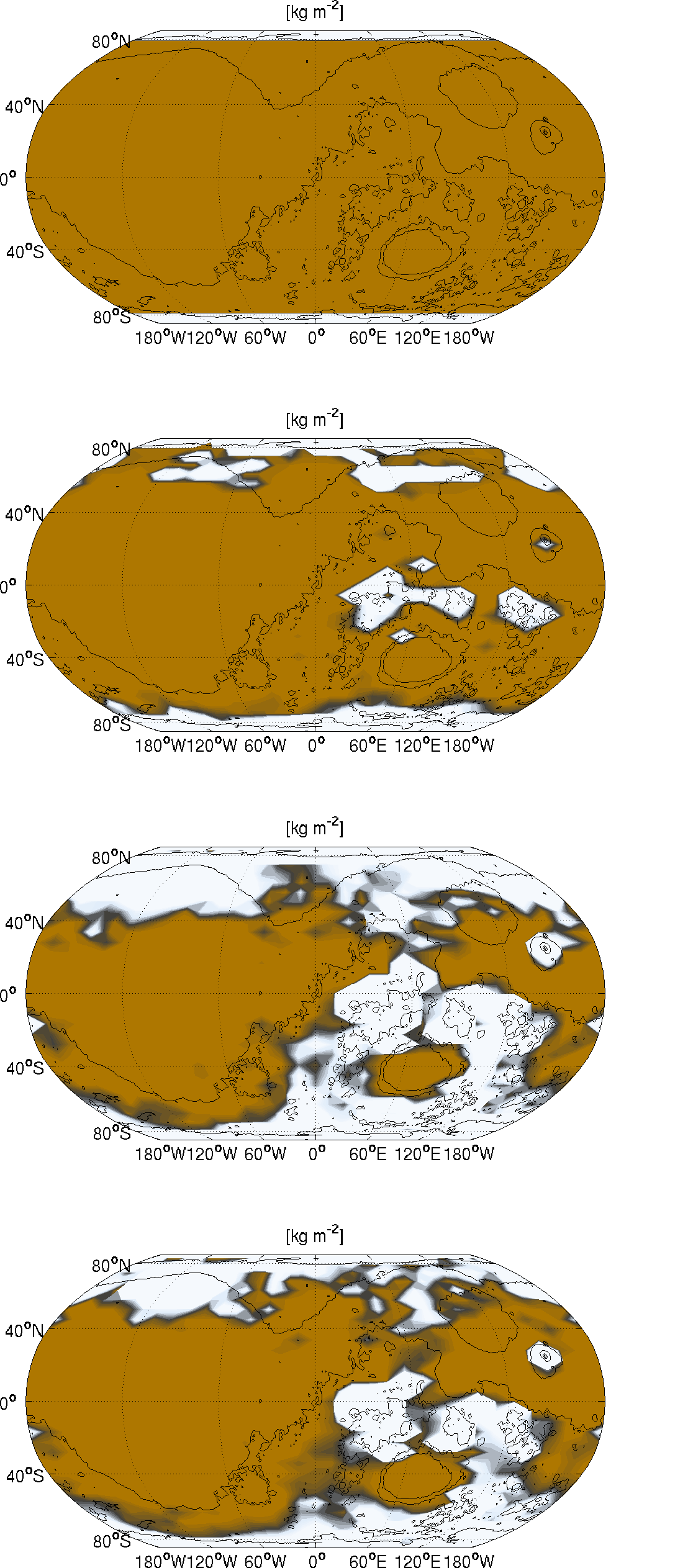}}
		{\includegraphics[width=2.25in]{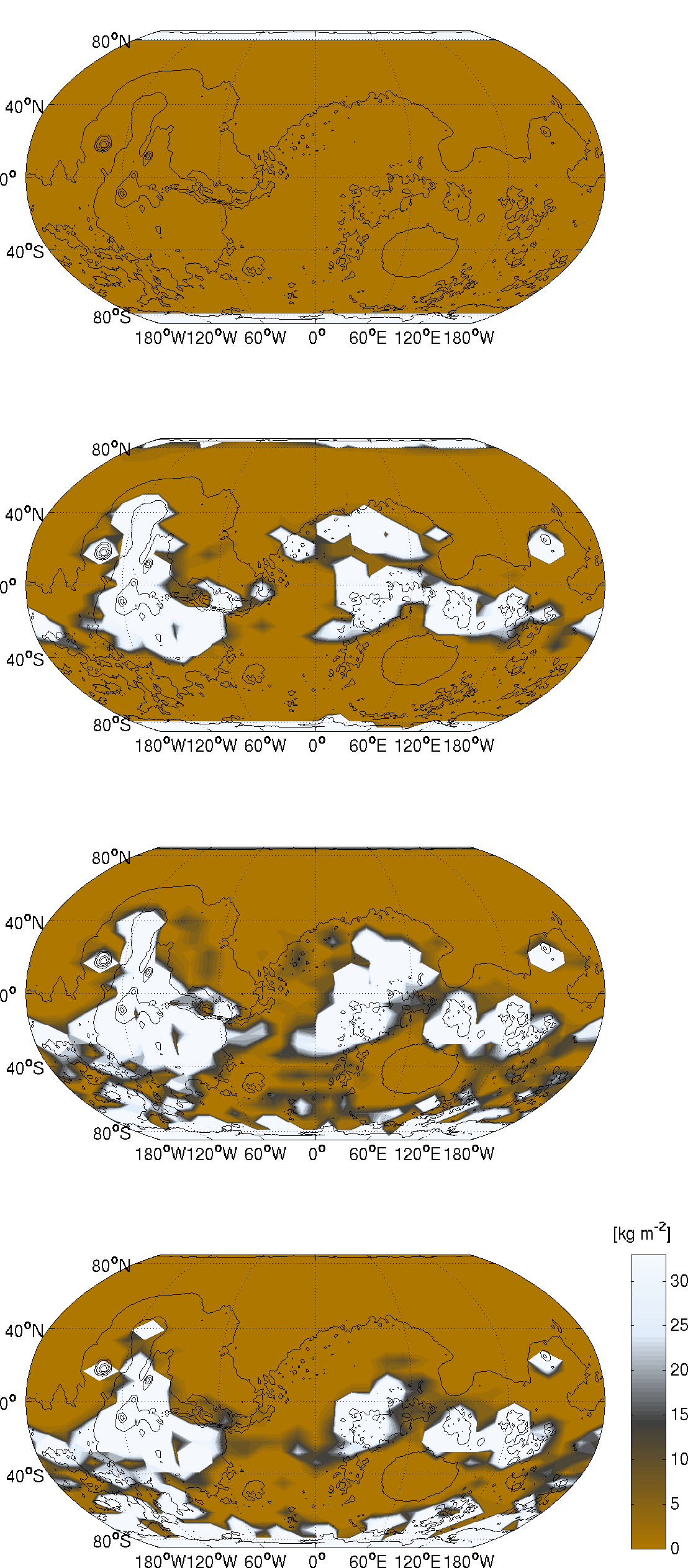}}
	\end{center}
	\caption{Same as Figure \ref{fig:iceevolplots1} (right column) except with modified topography (left) and with obliquity increased to 45$^\circ$ (right).}
	\label{fig:iceevolplots2}
 \end{figure}

\begin{figure}[h]
	\begin{center}
		{\includegraphics[width=2.25in]{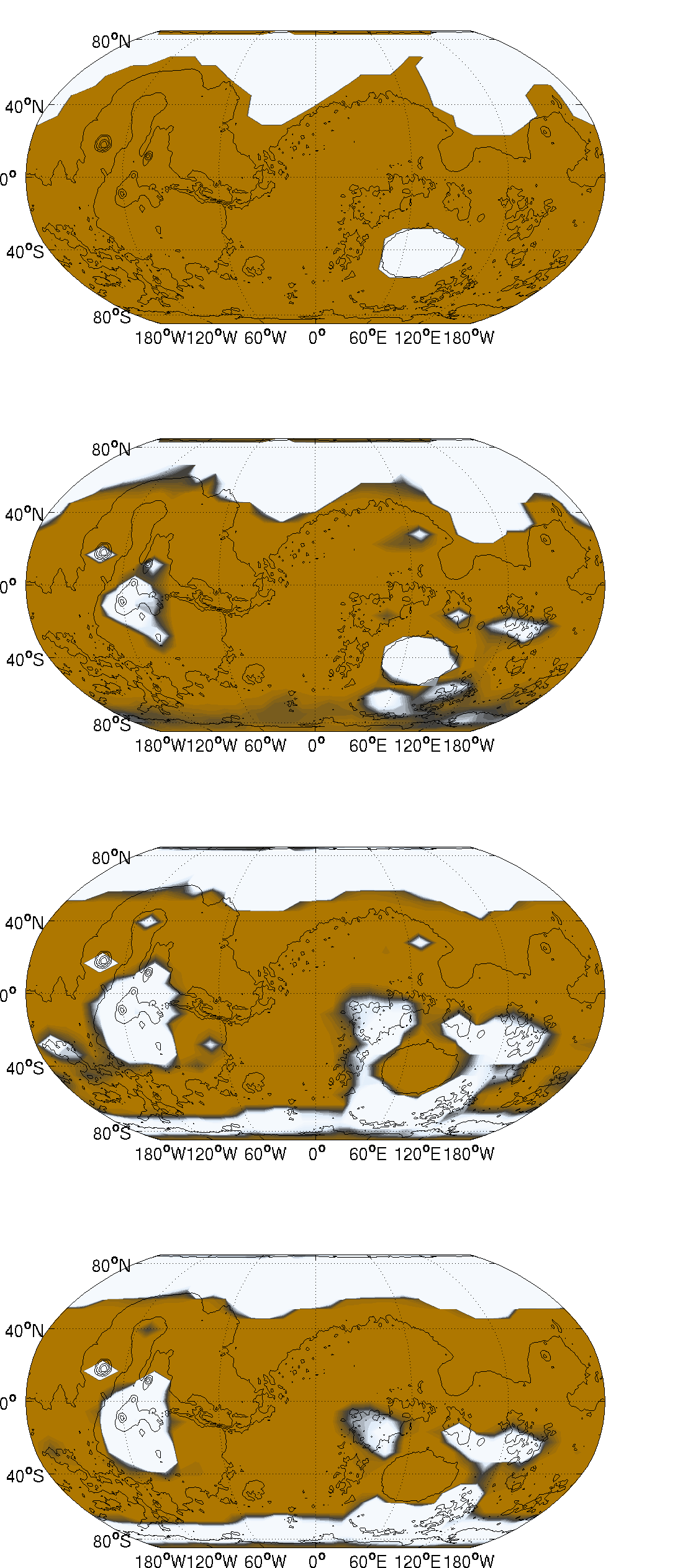}}
		{\includegraphics[width=2.25in]{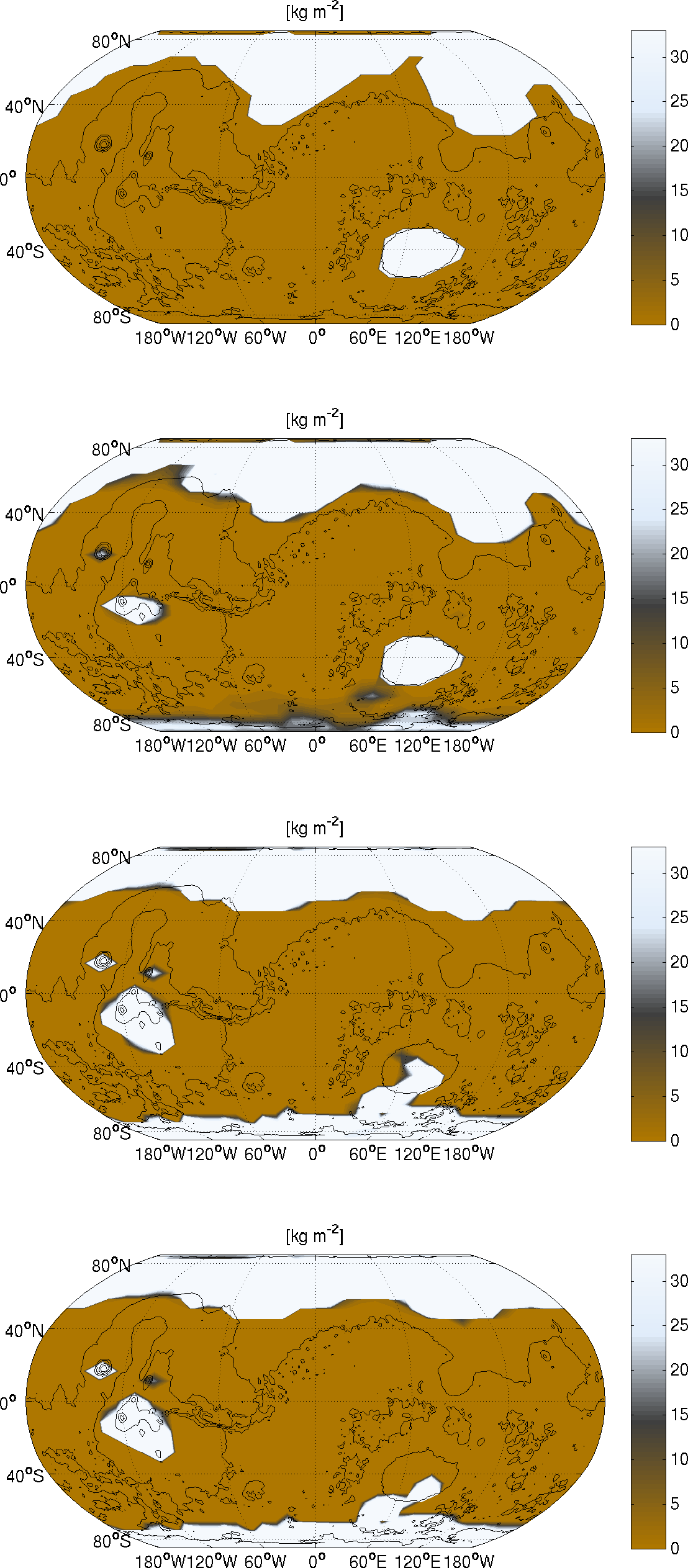}}
	\end{center}
	\caption{Same as Figure \ref{fig:iceevolplots1} (left column) except for surface pressure of 0.04 bar (left) and 0.008 bar (right).}
	\label{fig:iceevolplots3}
\end{figure}

Figures \ref{fig:iceevolplots1}-\ref{fig:iceevolplots3} show the evolution of the surface H$_2$O ice distribution for several simulations with different initial conditions and topography.  Snapshots of the ice are given at the start of the 1\textsuperscript{st}, 4\textsuperscript{th}, 20\textsuperscript{th} and 40\textsuperscript{th} years of simulation, with the iteration algorithm applied as described in Section \ref{subsec:icealgol}. Simulations were run for a range of pressures; here we focus on results at 0.008, 0.04 and 1 bar only.

Figure \ref{fig:iceevolplots1} shows two simulations with identical climate parameters (1~bar pressure, $\phi=25^\circ$) but different initial surface distributions of water ice (`icy lowlands' and `icy poles' for left and right columns, respectively). As can be seen, the two cases evolve towards similar equilibrium states, with ice present at both poles and in the highest-altitude regions across the planet (the Tharsis bulge, Olympus and Elysium Mons and the Noachian terrain around Hellas basin). The presence of significant amounts of ice over the equatorial regions means an increased planetary albedo. As a result of this and the reduced relative humidity due to localisation of surface H$_2$O sources, the mean surface temperature in the final year ($\sim$233 K) is several degrees below that of the equivalent 100\% humidity simulation. Figure \ref{fig:iceevolavged} shows the longitudinally and yearly averaged surface ice as a function of time and latitude for the same two simulations. The plots indicate that even at the end of the simulation, the ice density around the equator and in the south was still increasing slightly, mainly at the expense of deposits around 40 $^\circ$S. Analysis of the atmospheric dynamics (see next section) suggested that low relative humidity associated with convergence of the annual mean meridional circulation at mid-latitudes was responsible for this long-term effect.

In Figure \ref{fig:iceevolplots2} (left column), the same results are plotted for a 1~bar, $\phi=25^\circ$ case with modified surface topography as described by equation (\ref{eq:nuketharsis}). Here evolution is similar to that in the standard cases, except that the correlation between ice deposits and the distribution of Noachian-era valley networks \citep[see e.g. ][]{Fassett2008a,Hynek2010} is even clearer because the Tharsis bulge is absent. When the obliquity is increased, results are also broadly similar (see $\phi=45^\circ$ case; right column of Figure \ref{fig:iceevolplots2}), except that ice disappears from the northern pole entirely due to the increased insolation there. The effects of obliquity on ice transport has been discussed in the context of the present-day Martian atmosphere by \cite{Forget2006} and \cite{Madeleine2009}. In these dense-atmosphere simulations, the combination of adiabatic heating and increased insolation in the northern plains at high obliquities makes them the warmest regions of the planet, with mean temperatures of $\sim$ 235 K even for latitudes north of 5 $^\circ$N.

As a test of the ice evolution algorithm, we also performed simulations at 0.008 and 0.04 bar pressure (Figure \ref{fig:iceevolplots3}). Converging on a solution proved challenging in these cases, as the rate of H$_2$O ice transport was extremely slow, and permanent CO$_2$ ice caps formed at the poles as on present-day Mars. At 0.008 bar, the total atmospheric mass was so low that after several years the simulations approached a seasonally varying CO$_2$ vapour-pressure equilibrium. At 0.04 bar, however, atmospheric pressure continued to decrease slowly throughout the simulation.

As a result of the large H$_2$O ice evolution timescales, the simulations did not reach an exact equilibrium state even after the full 40 years of simulation time. Nonetheless, Figure \ref{fig:iceevolplots3} clearly shows the differences from the 1-bar case: H$_2$O ice is present in large quantities at the poles, with a smaller deposit over the Tharsis bulge. Analysis of the long-term ice trends in these cases (not shown) indicated that the southern polar caps were still slowly growing even at the end of the simulations.

The differences in ice migration between the low and high pressure simulation are easily understood through the adiabatic effect described earlier and shown in Figure \ref{fig:tsurf2D}. In all cases, ice tends to migrate towards the coldest regions on the surface. At 0.008 bar (Figure \ref{fig:tsurf2D}, left-hand column), these regions correlate almost exactly with latitude, while at 1 bar (Figure \ref{fig:tsurf2D}, right-hand column), the correlation is primarily altitude-dependent.

\subsubsection{Atmospheric dynamics and clouds}
\begin{figure}[h]
	\begin{center}
		{\includegraphics[height=1.3in]{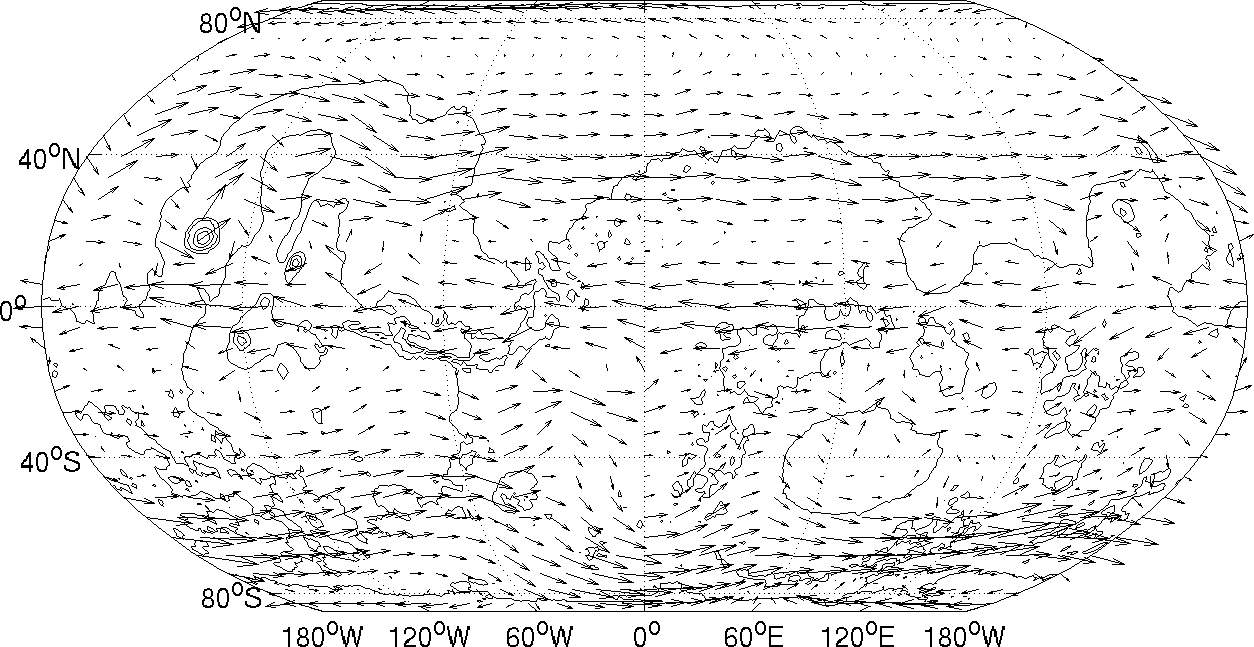}}
    	{\includegraphics[height=1.5in]{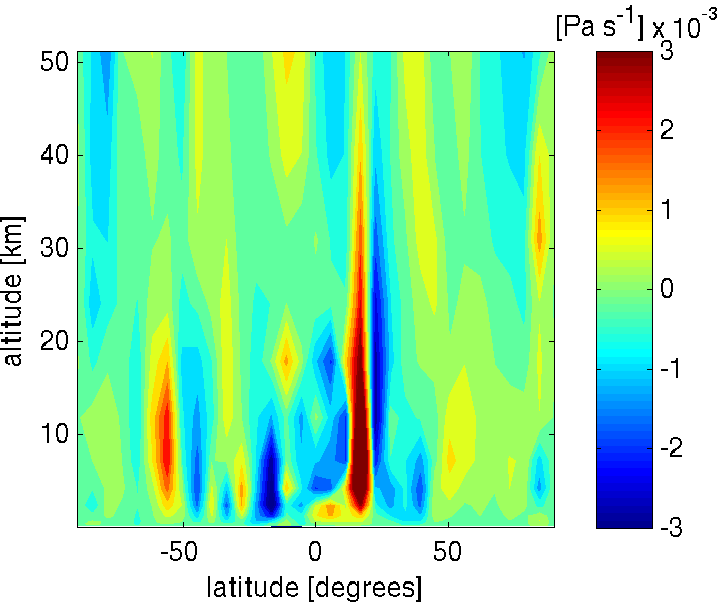}}\
	\end{center}
	\caption{(left) Annual mean horizontal wind in the middle atmosphere (at approx. 500~hPa) and (right) annual and zonal mean vertical velocity (right), for the 1-bar 25$^\circ$ obliquity case after 40 simulation years.}
	\label{fig:weather}
\end{figure}

\begin{figure}[h]
	\begin{center}
		{\includegraphics[height=1.3in]{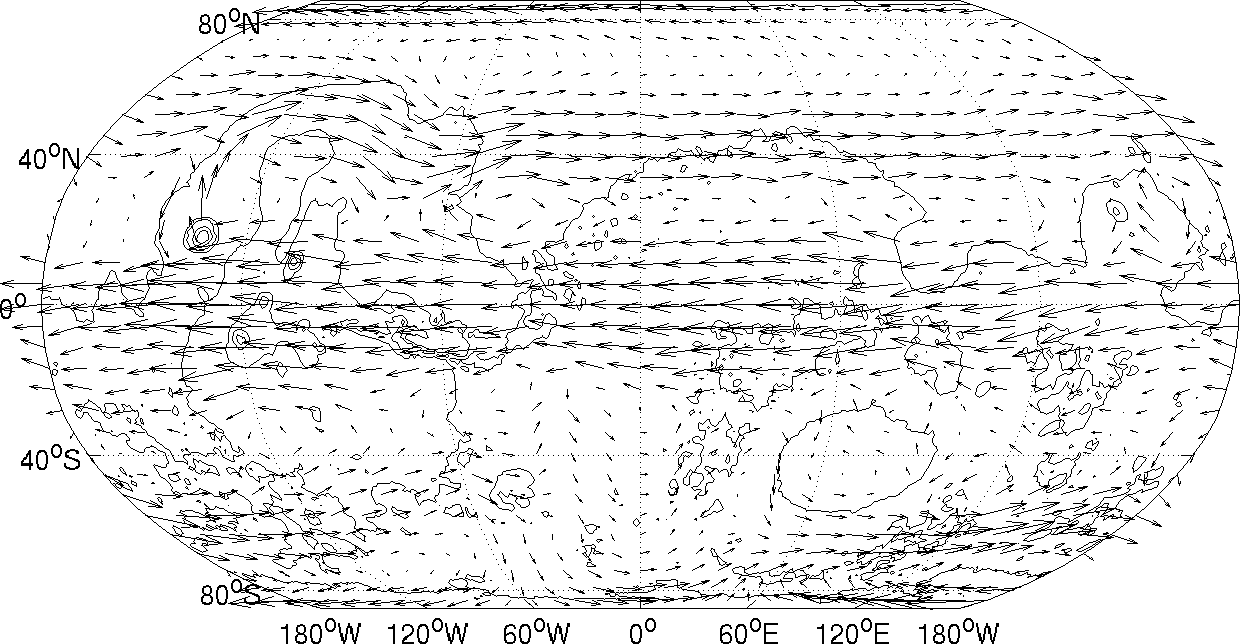}}
		{\includegraphics[height=1.5in]{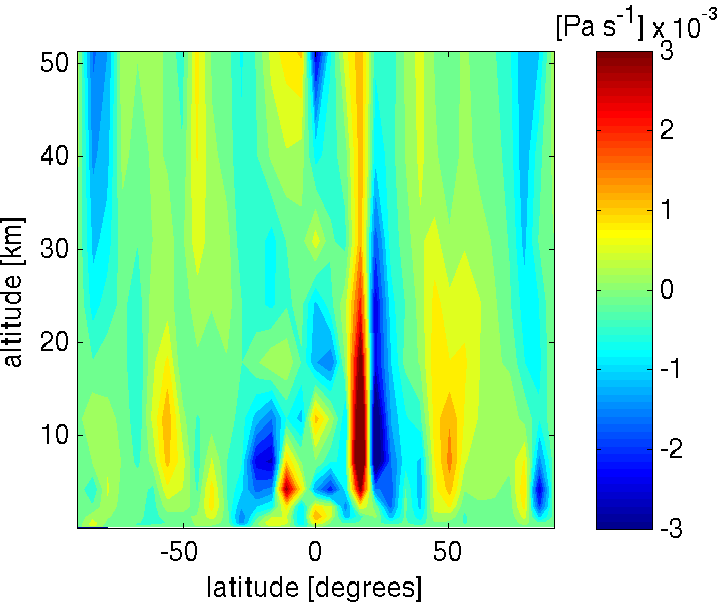}}
	\end{center}
	\caption{Same as Figure \ref{fig:weather}, but for 45$^\circ$ obliquity.}
	\label{fig:weather2}
\end{figure}

Although the primary factor in the H$_2$O ice distribution is the annual mean surface temperature distribution, the final steady state was modulated by the effects of the global circulation. Figures \ref{fig:weather} and \ref{fig:weather2} show the annual mean horizontal velocity at the 9th vertical level (approx. 500~hPa; left) and the annual and zonal mean vertical velocity $\omega$ (right) for the final year of the 1-bar simulation with obliquity 25$^\circ$ and 45$^\circ$, respectively. As can be seen, the annual mean circulation consists of an equatorial westward jet, with a transition to eastward jets above absolute latitudes of 40$^\circ$. The associated vertical velocities are asymmetric as a result of the planet's north/south topographic dichotomy, with intense vertical shear near 20$^\circ$N in both cases. Nonetheless, wide regions of downwelling (positive $\omega$) are apparent around 50$^\circ$N/S. In analogy with the latitudinal relative humidity variations due to the meridional overturning circulation on Earth, these features explain the tendency of ice to migrate away from mid-latitudes, as apparent from Figures \ref{fig:iceevolplots1}-\ref{fig:iceevolplots2} (described in the last section).

\begin{figure}[h]
	\begin{center}
		{\includegraphics[height=1.5in]{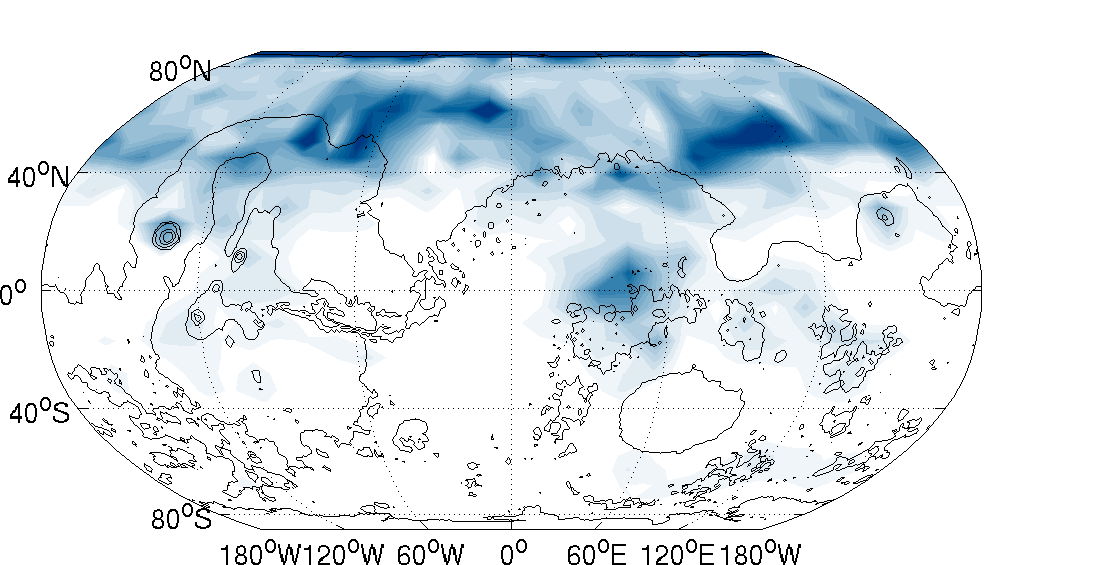}}
		{\includegraphics[height=1.5in]{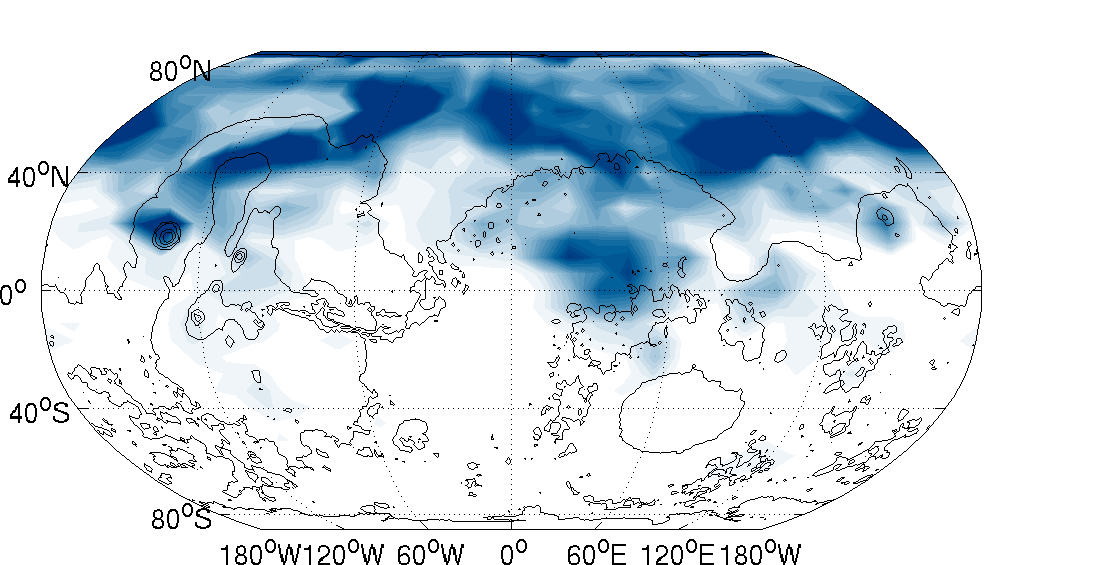}}
		{\includegraphics[height=1.5in]{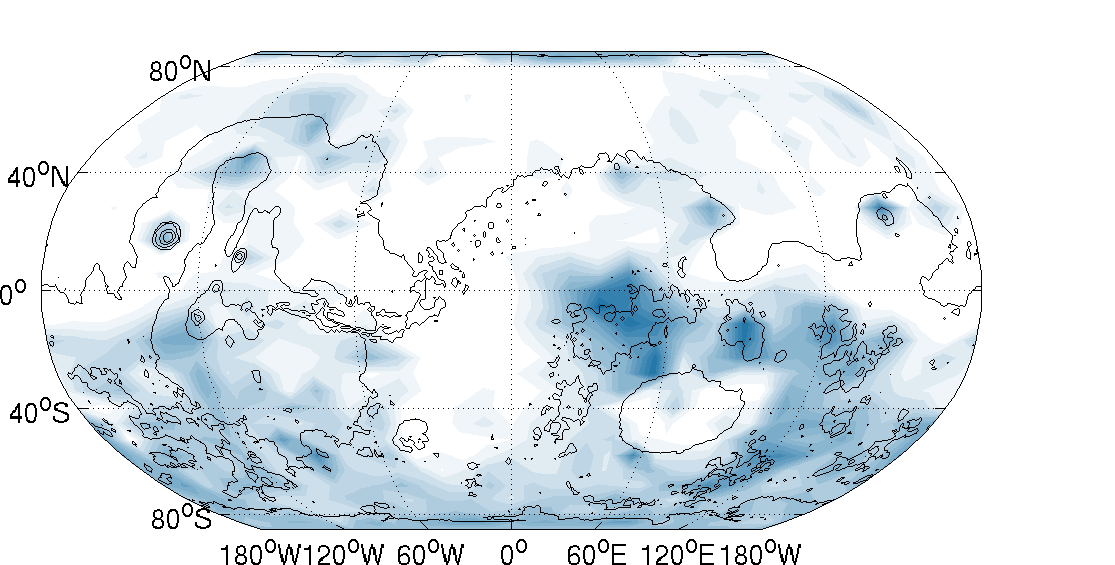}}
	{\includegraphics[height=1.5in]{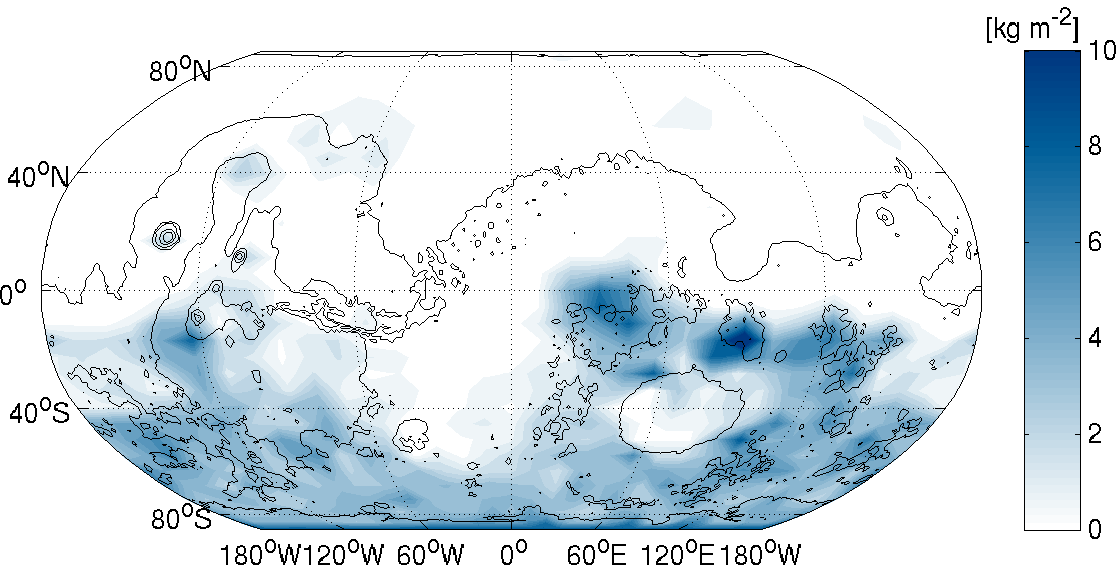}}
	\end{center}
	\caption{Total precipitation (snowfall) in each season for $L_s=0^\circ$ to $90^\circ$, $90^\circ$ to $180^\circ$,  $180^\circ$ to $270^\circ$ and $270^\circ$ to $360^\circ$ (in descending order), for the 1-bar 25$^\circ$ obliquity case after 40 simulation years.}
	\label{fig:precip}
\end{figure}

Figure \ref{fig:precip} shows the total precipitation in each season for the same simulation (obliquity 25$^\circ$). It highlights the dramatic annual variations in the meteorology. From  $L_s = 0^\circ$ to $180^\circ$ (top two maps), relatively intense precipitation (snowfall) occurred in the northernmost regions due to the increased surface temperatures there. In contrast, from  $L_s = 180^\circ$ to $360^\circ$ (bottom two maps) lighter snowfall occurred in the south, particularly in the high Noachian terrain around Hellas basin where the valley network density is greatest.

\begin{figure}[h]
	\begin{center}
		{\includegraphics[height=3.5in]{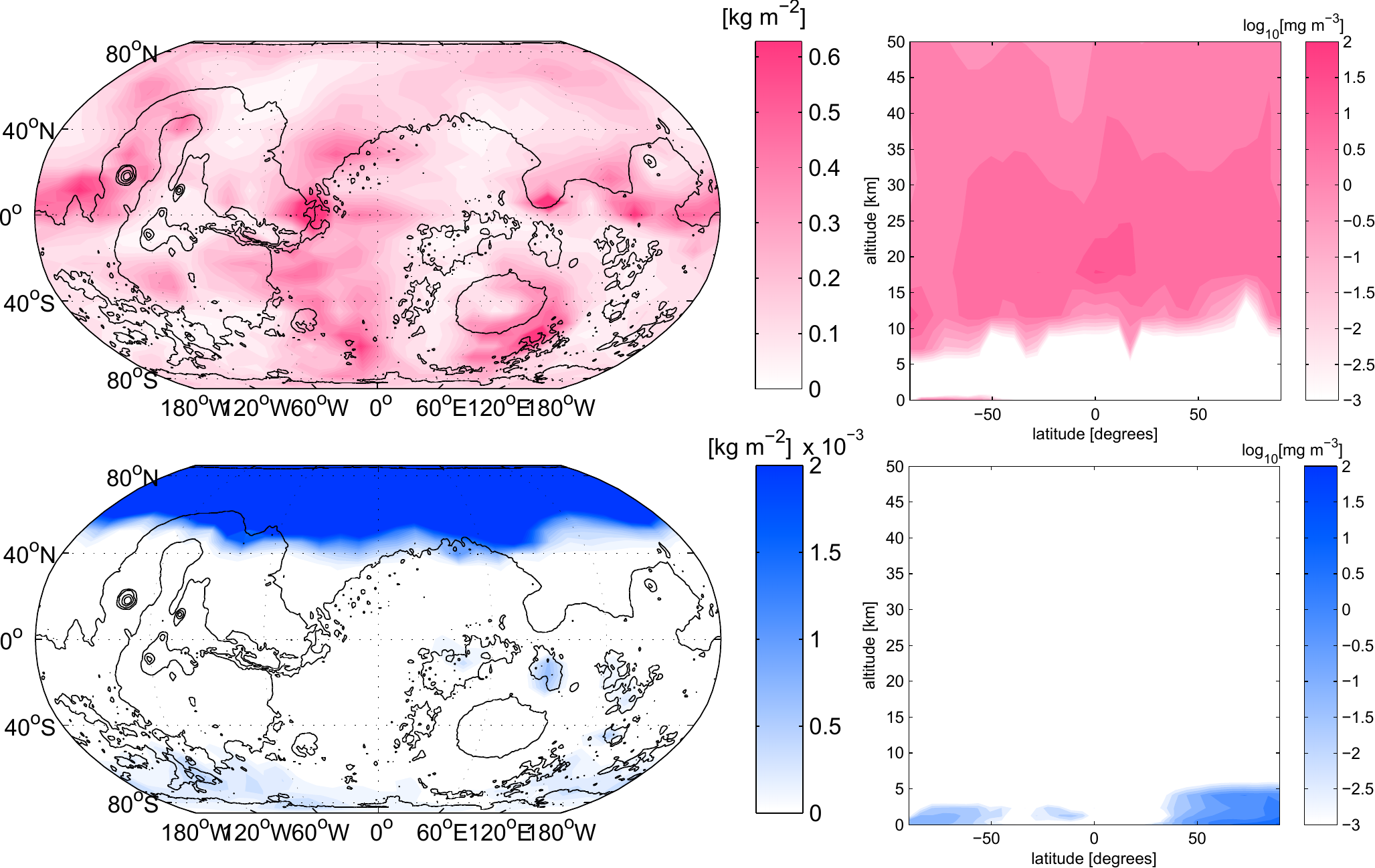}}
	\end{center}
	\caption{Annual mean column amounts (left) and annual and zonal mean mass mixing ratios (right) of CO$_2$ (red) and H$_2$O (blue) cloud condensate for the 1-bar 25$^\circ$ obliquity case  after 40 simulation years.}
	\label{fig:clouds}
\end{figure}

Figure \ref{fig:clouds} shows annual mean column amounts (left) and annual and zonal mean mass mixing ratios (right) of cloud condensate (CO$_2$ and H$_2$O) for the same simulation. As can be seen, CO$_2$ clouds form at high altitudes and vary relatively little with latitude. H$_2$O clouds form much lower in locations that are dependent on the surface water sources. Because of the Martian north-south topographic dichotomy, H$_2$O cloud content was much greater in the northern hemisphere, where temperatures were typically warmer in the low atmosphere. We found typical H$_2$O cloud particle sizes of 2-10 $\mu$m in our simulations, although these values were dependent on our choice of cloud condensation nuclei parameter $[CCN]$ (set to $10^5$ kg$^{-1}$ for all the main simulations described here; see Section \ref{sec:method} and Table \ref{tab:params}).

To assess the radiative impact of the H$_2$O clouds, we performed some simulations where the H$_2$O cloud opacity was set to zero after climate equilibrium had been achieved. Because global mean temperatures were low in our simulations even at one bar CO$_2$ pressure and the planetary albedo was already substantially modified by higher-altitude CO$_2$ clouds, only small changes (less than 5~K) in the climate were observed. As we used a simplified precipitation scheme and assumed that the water clouds had 100\% coverage in each horizonal grid cell of the model, we may have somewhat inaccurately represented their radiative effects. However, mean surface temperatures were already well below 273 K in our saturated water vapour simulations of Section \ref{subsec:thermalresults} at the same pressures, so this is unlikely to have qualitatively affected our results. For a detailed discussion of the coverage and microphysics of CO$_2$ clouds in the simulations, refer to \cite{Forget2012}.

To assess the uncertainties in our H$_2$O cloud parameterisation further, we also performed some tests where we varied the number of condensation nuclei $[CCN]$ for H$_2$O and the precipitation threshold $l_0$ (Table \ref{tab:sens}). In all tests, we started from the 1~bar, $\phi=25^\circ$ simulations in equilibrium and ran the model for 8 Mars years without ice evolution. We found the effects to be modest; across the range of parameters studied, the total variations in mean surface temperature were under 1~K. The dominating influence of high CO$_2$ clouds on atmospheric radiative transfer at 1~bar pressure was the most likely cause of this low sensitivity.

When the precipitation threshold was removed altogether, H$_2$O clouds became much thicker optically, leading to larger climate differences. In these tests, we observed significant transients in surface temperature, with increases to 250-260~K after one to two years before a slow decline to 200-215~K, after which time permanent CO$_2$ glaciation usually began to occur. This long-term cooling effect was caused by increased reflection of solar radiation by an extremely thick H$_2$O cloud layer, as evidenced by the increased cloud condensate column density and planetary albedo values (Table \ref{tab:sens}). Particle coagulation is a fundamental physical process in water liquid/ice clouds, and the mean atmospheric densities of condensed H$_2$O in these simulations appeared extremely unrealistic when compared to e.g., estimated values for the Earth under snowball glaciation conditions \citep{Abbot2012}. We therefore did not regard these latter tests to be physically robust, and used the threshold scheme for all simulations with a water cycle presented here. Nonetheless, it is clear that further research in this area in future using more sophisticated cloud schemes would be useful.

\begin{table}[h]
\centering
\caption{Sensitivity of the results to H$_2$O cloud microphysical parameters. From left to right the columns show the prescribed values of $[CCN]$ and $l_0$, and the simulated global annual mean surface temperature, planetary albedo, H$_2$O cloud column density, and H$_2$O vapour column density after 10~years simulation time. In all tests $[CCN]$ was kept constant at $1\times10^5$~kg$\slash$kg for the CO$_2$ clouds.}
\begin{tabular}{cc|cccc}
\hline
$[CCN]$ [kg$\slash$kg] & $l_0$ [kg$\slash$kg] & $\overline T_s$ [K] & $A_p$ & $\overline q_{H_2O \mbox{ cond.}}$ [kg m$^{-2}$]   & $\overline q_{H_2O \mbox{ vap.}}$ [kg m$^{-2}$] \\
\hline
$1\times10^4$ & 0.001 & 233.2     & 0.45 & $6.5\times10^{-4}$ & 0.069 \\
$1\times10^5$ & 0.001 & 232.8     & 0.45 & $4.7\times10^{-4}$ & 0.065 \\
$1\times10^6$ & 0.001 & 233.5     & 0.45 & $5.5\times10^{-4}$ & 0.070 \\
$1\times10^5$ & 0.01  & 233.2     & 0.45 & $7.2\times10^{-4}$ & 0.068 \\
$1\times10^4$ & $\infty$ & 214.0 & 0.77 & $11.1$ & 0.044 \\
$1\times10^5$ & $\infty$ & 212.5 & 0.76 & $6.37$ & 0.036 \\
$1\times10^6$ & $\infty$ & 199.8 & 0.79 & $1.17$ & 0.056 \\
\hline
\end{tabular}\label{tab:sens}
\end{table}

\subsubsection{Transient melting}\label{subsec:transmelt}
\begin{figure}[h]
	\begin{center}
		{\includegraphics[width=3in]{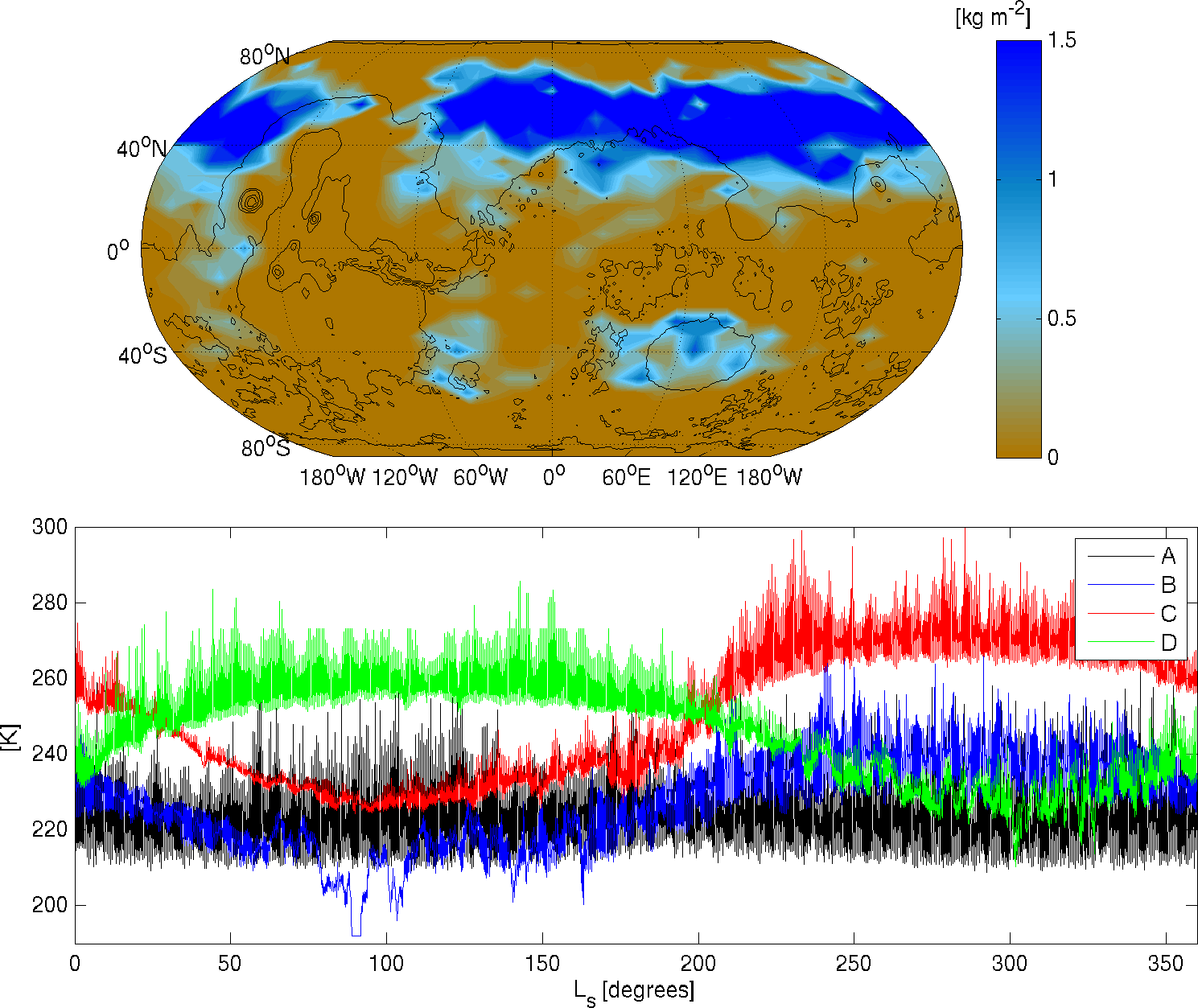}}
	\end{center}
	\caption{(top) Maximum surface liquid water in one year after 40 years simulation time, for the 1-bar simulation with obliquity 25$^\circ$ shown in Figure \ref{fig:iceevolplots1}. (bottom) Surface temperature vs. time for the same simulation, for the four locations displayed in Figure \ref{fig:TLOCS}.}
	\label{fig:transmelt}
\end{figure}

Figure \ref{fig:transmelt} (top) shows a contour plot of the annual maximum surface liquid H$_2$O at each gridpoint after 40 years simulation time for the case shown in Figure \ref{fig:iceevolplots1} (1 bar, 25$^\circ$ obliquity). Given the relatively low spatial resolution of our model and the lack of accurate parameterisations for important sub-gridscale processes (slope effects, small-scale convection etc.), these results can only give an approximate guide to local melting under the simulated global climate. Nonetheless, it is clear that liquid water appears transiently in some amounts across the planet. The increased melting in the northern hemisphere can  be explained by the higher temperatures there due to the same adiabatic effect responsible for the migration of ice to the southern highlands.

\begin{figure}[h]
	\begin{center}
		{\includegraphics[width=3in]{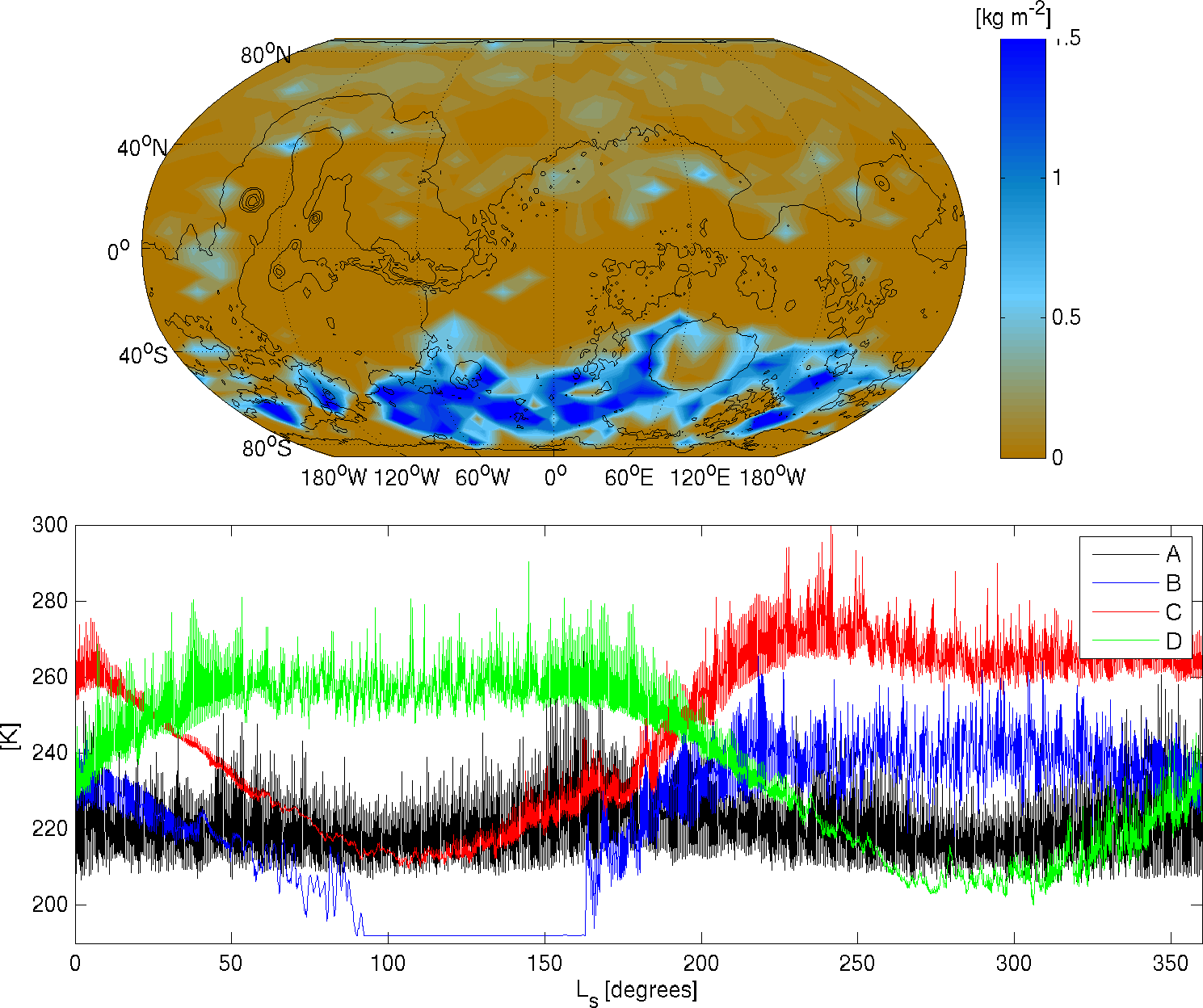}}
	\end{center}
	\caption{Same as Figure \ref{fig:transmelt}, except for obliquity 45$^\circ$.}
	\label{fig:transmelt2}
\end{figure}

Figure \ref{fig:transmelt} (bottom) shows surface temperature against time for the four locations shown in Figure \ref{fig:TLOCS}. The dramatic difference in temperatures between the Tharsis bulge (A) and the bottom of Hellas basin (C) is clear, along with the increased seasonal variation away from the equator (C and D). The transient melting occurring at location D (Utopia Planitia) is clear from the peak of surface temperatures at 273~K there between $L_s \sim 70$ and 120$^\circ$. Further heating does not occur in this period because some surface H$_2$O ice is still present. In Figure \ref{fig:transmelt2}, the same results are plotted for the $\phi = 45^\circ$ case. There, the differences in insolation and absence of ice in the north mean that the great majority of melting events occur south of the equator. As can be seen from the surface temperature plot (bottom), conditions at the equator become cold enough for seasonal CO$_2$ ice deposits to form on the Tharsis bulge (location A; blue line).

\begin{figure}[h]
	\begin{center}
		{\includegraphics[width=3in]{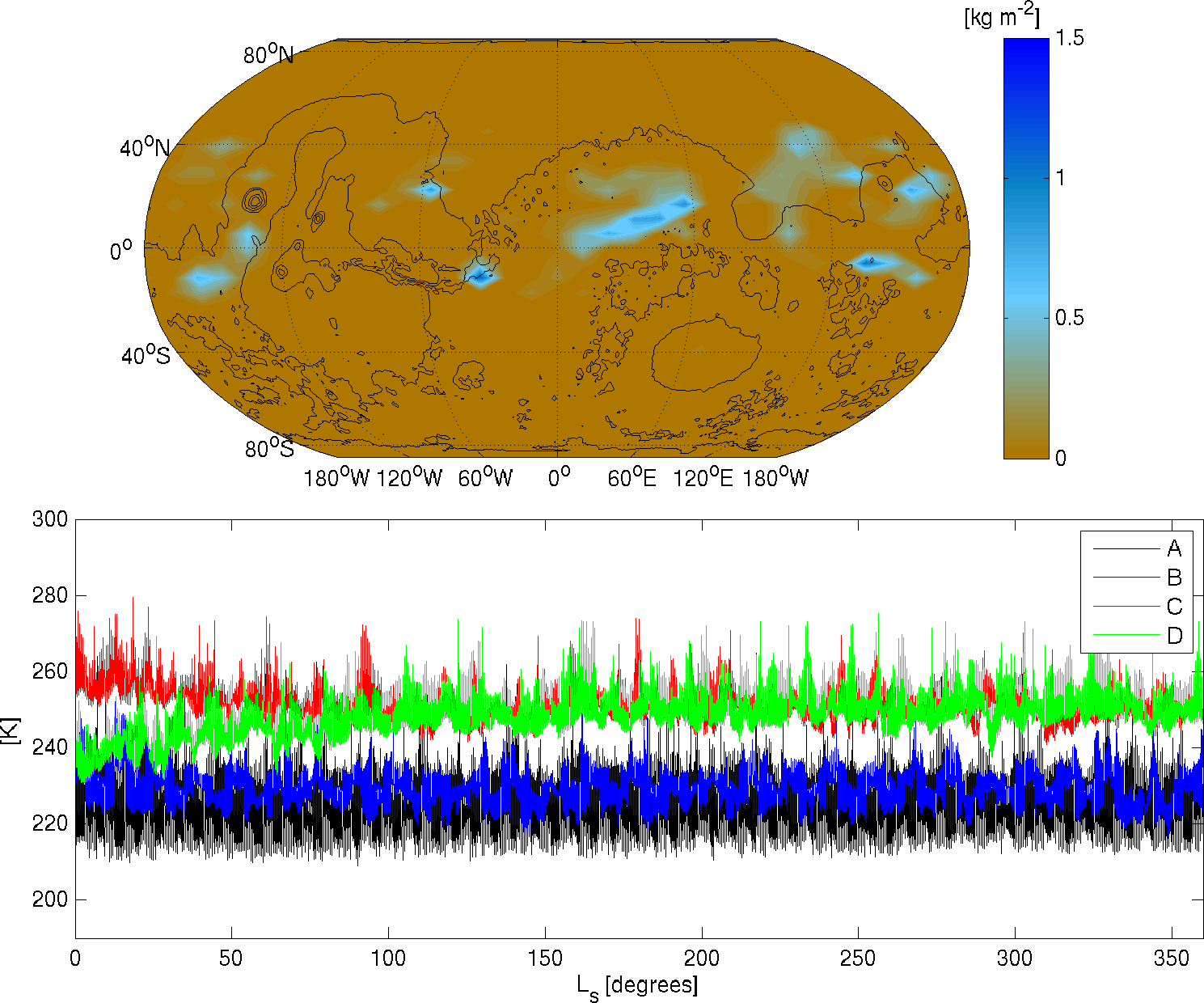}}
	\end{center}
	\caption{Same as Figure \ref{fig:transmelt}, except with obliquity set to 0$^\circ$ after 40 years simulation time.}
	\label{fig:transmelt3}
\end{figure}

\begin{figure}[h]
	\begin{center}
		{\includegraphics[width=3in]{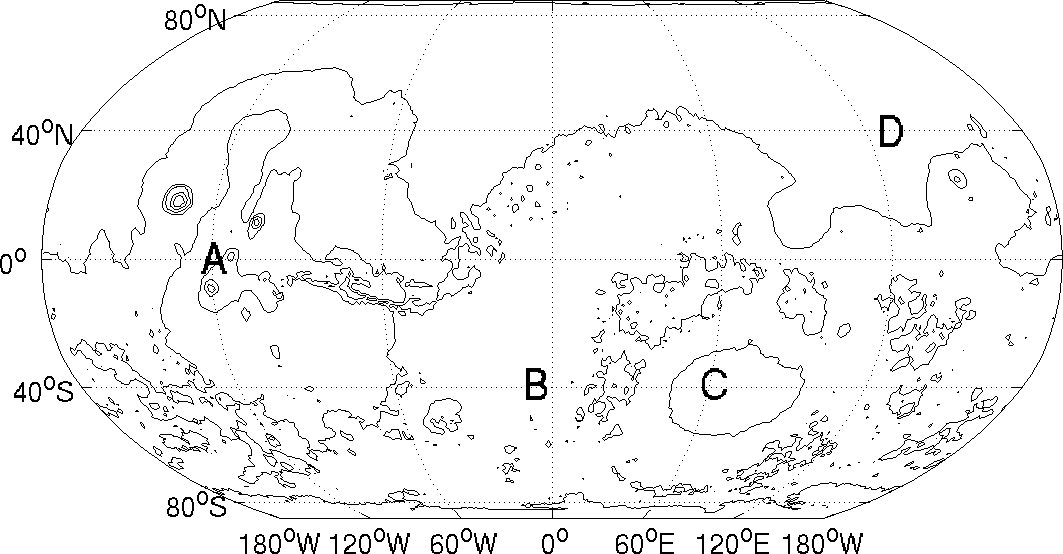}}
	\end{center}
	\caption{Locations for the four temperature time series displayed in Figs. \ref{fig:transmelt}-\ref{fig:transmelt3}.}
	\label{fig:TLOCS}
\end{figure}

It is well known that Mars' obliquity evolves chaotically, and during the planet's history, relatively rapid changes in a range from 0 to 70$^\circ$ are likely to have occurred many times \citep{Laskar2004}. With this in mind, we investigated the effects of decreasing obliquity to zero, but keeping the end-state surface ice distribution of Figure \ref{fig:iceevolplots1}. An obliquity of zero maximises insolation at the equator, which might be expected to maximise melting in the valley network region. However, for this case we found that while mean equatorial temperatures were slightly higher, the lack of seasonal variation reduced annual maximum temperatures in most locations. Hence there was a net decrease in the annual maximum surface liquid water (Figure \ref{fig:transmelt3}).

\begin{figure}[h]
	\begin{center}
		{\includegraphics[width=2.75in]{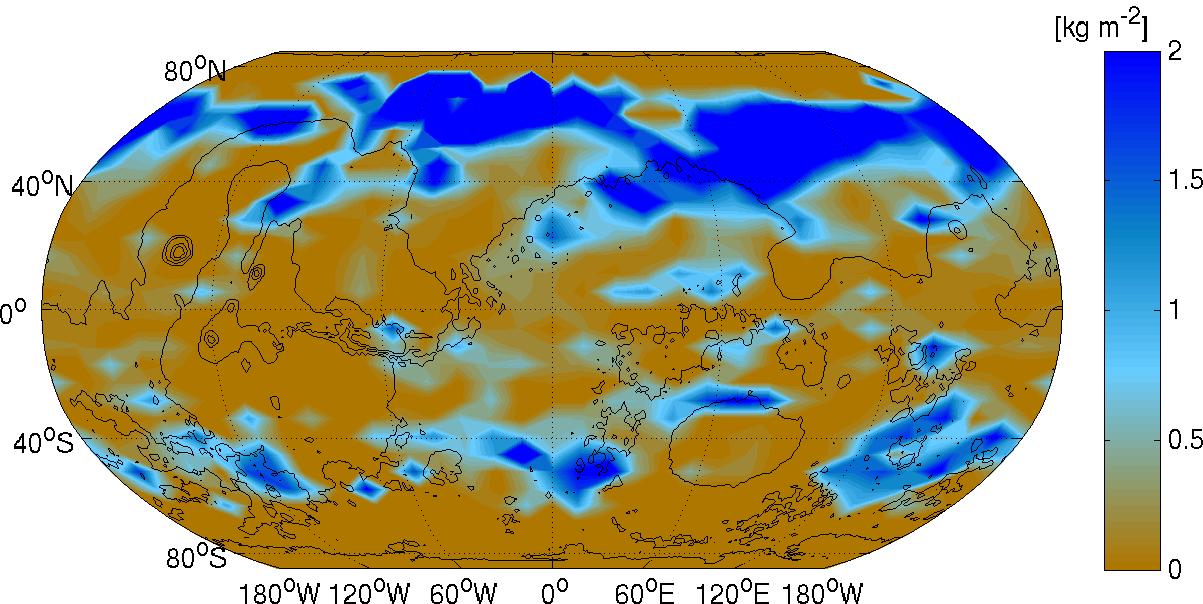}}
		{\includegraphics[width=2.75in]{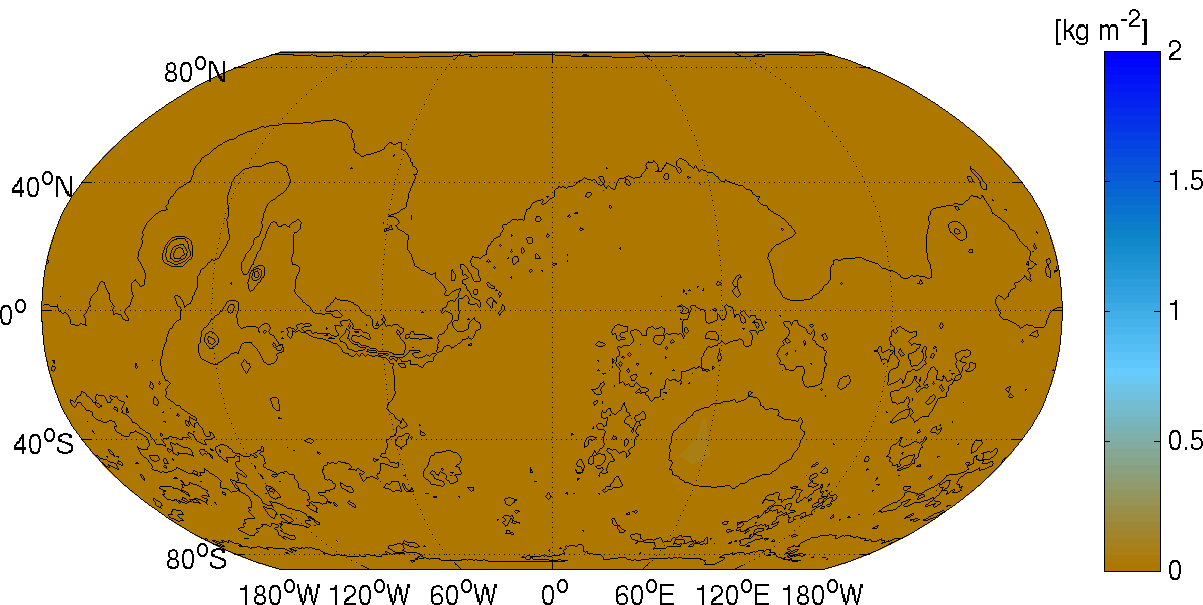}}
	\end{center}
	\caption{Maximum surface liquid water in one year after 40 years simulation time, for the 1-bar simulation with obliquity 25$^\circ$ shown in Figure \ref{fig:iceevolplots1}, after (left) change of ice surface albedo from 0.5 to 0.3 and (right) increase of ice surface thermal inertia from 250 to 1000 J~m$^{-2}$~s$^{-1\slash 2}$~K$^{-1}$.}
	\label{fig:transmelt4}
\end{figure}

To test the sensitivity of the amount of transient melting to our assumed parameters, we also performed some simulations where we varied the H$_2$O ice surface albedo and thermal inertia. Starting from the 1 bar, 25$^\circ$ obliquity simulation just described, we a) reduced ice albedo from 0.5 to 0.3 and b) increased  the surface ice thermal inertia from $\mathcal I =$ 250 to 1000 J~m$^{-2}$~s$^{-1\slash 2}$~K$^{-1}$ at all soil depths. In both cases all other parameters were kept constant. Figure \ref{fig:transmelt4} shows the results for these two tests in terms of the annual maximum surface liquid water.

Perhaps unsurprisingly, when the ice albedo was reduced melting occurred over a wider range of surface locations, and the total amount of melting in each year slightly increased. As a variety of processes may influence this parameter, including dust transport and volcanic ash emission \citep[e.g., ][]{Wilson2007,Kerber2011}, this may have interesting implications for future study. However, increasing the surface ice thermal inertia essentially shut down transient melting entirely (Figure \ref{fig:transmelt4}, right). As this calculation was performed assuming constant thermal inertia with depth, it clearly represents a lower limit on potential melting. Nonetheless, it indicates that surface seasonal (as opposed to basal) melting on early Mars under these conditions would almost certainly be limited to small regions on the edges of ice sheets only. To constrain the values of these parameters better, a more detailed microphysical / surface model that included the effects of sub-gridscale ice deposition, dust and possibly volcanism would be required.

\subsubsection{Southern polar ice and the Dorsa Argentea Formation}
In our simulations, we find that under a dense atmosphere, thick ice sheets form over Mars' southern pole. Ice migrates there preferentially from the north because of the same adiabatic cooling effect responsible for deposition over Tharsis and the equatorial highlands. The Dorsa Argentea Formation is an extensive volatile-rich south polar deposit, with an area that may be as great as 2~\% of the total Martian surface \citep{HeadPratt2001}. It is believed to have formed in the late Noachian to early Hesperian era \citep{Plaut1988}, when the atmosphere is likely to have still been thicker than it is today. Given that our simulations span a range of CO$_2$ pressures, it is interesting to compare the results with geological maps of this region.

\begin{figure}[h]
	\begin{center}
		{\includegraphics[width=5in]{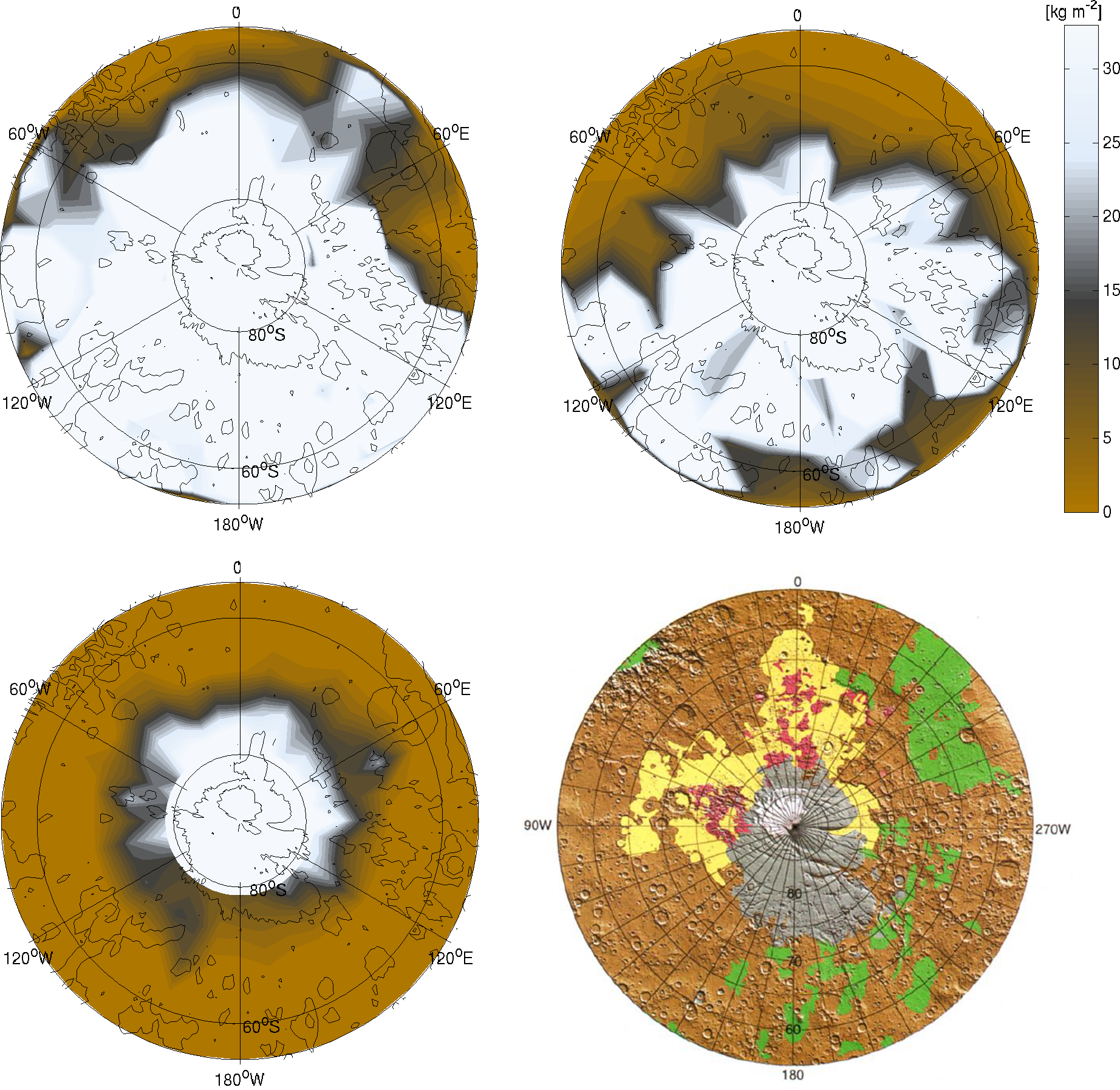}}
	\end{center}
	\caption{Plots of the annual mean water ice coverage on the southern pole after 40 years for mean surface pressure 1 bar and obliquity a) 25$^\circ$, b) 45$^\circ$ and mean surface pressure 0.2 bar and obliquity 25$^\circ$. d) For comparison, a plot of the Dorsa Argentea Formation from \cite{HeadPratt2001} is also shown.}
	\label{fig:dorsaeargentea}
\end{figure}

As an example, Figure \ref{fig:dorsaeargentea} shows surface ice in simulations at 1 bar (25 and 45$^\circ$ obliquity) and 0.2 bar (25$^\circ$ obliquity), with a map plotted alongside that shows the extent of the main geological features \citep{HeadPratt2001}. In Figure \ref{fig:dorsaeargentea}d, yellow and purple represent the Dorsa Argentea Formation lying on top of the ancient heavily cratered terrain (brown) and below the current (Late Amazonian) polar cap (gray and white). As can be seen, there is a south polar H$_2$O cap in all cases, although its latitudinal extent is greatest at 1 bar (Figure \ref{fig:dorsaeargentea}a). At current obliquity and between 0.2 and 1 bar (Figure \ref{fig:dorsaeargentea}a, c), the accumulation generally covers the area of the Dorsa Argentea Formation, and also extends toward 180 degrees longitude in the 0.2 bar case, a direction that coincides with the southernmost development of valley networks \citep{Fassett2008a,Hynek2010}. Further simulations of ice formation in this region for moderately dense CO$_2$ atmospheres  would be an interesting topic for future research. In particular, it could be interesting to study southern polar ice evolution using a zoomed grid or mesoscale model, with possible inclusion of modifying effects due to the dust cycle.

\section{Discussion}\label{sec:discuss}

In contrast to the `warm, wet' early Mars envisaged in many previous studies, our simulations depict a cold, icy planet where even transient diurnal melting of water in the highlands can only occur under extremely favourable circumstances. As abundant quantities of liquid water clearly did flow on early Mars, other processes besides those included in our model must therefore have been active at the time. Many cold-climate mechanisms for Martian erosion in the Noachian have been put forward previously, including aquifer recharge by hydrothermal convection \citep{Squyres1994} and flow at sub-zero temperatures by acidic brines \citep{Gaidos2003,Fairen2010}. In the following subsections, we focus on three particularly debated processes: heating by impacts, volcanism, and basal melting of ice sheets.

\subsection{Meteorite impacts}
\label{subsec:impacts}

As noted in Section \ref{sec:intro}, impacts have already been proposed by \cite{Segura2002,Segura2008} as the primary cause of the valley networks via the formation of transient steam atmospheres. However, one of the key Segura et al. arguments, namely that the rainfall from these atmospheres post-impact is sufficient to reproduce the observed erosion, has been strongly criticized by other authors \citep{Barnhart2009}. Nonetheless, the numerous impacts that occurred during the Noachian should have caused local melting of surface ice if glaciers were present. Such a mechanism has been suggested to explain fluvial landforms on fresh Martian impact ejecta \citep{Morgan2009,Mangold2012}. Under the higher atmospheric pressure and greater impactor flux of the Noachian, it could have caused much more extensive fluvial erosion. However, for impact-induced melting to be a viable explanation for the valley networks, some mechanism to preferentially transport water to the southern highlands must be invoked.

\begin{figure}[h]
	\begin{center}
		{\includegraphics[width=3in]{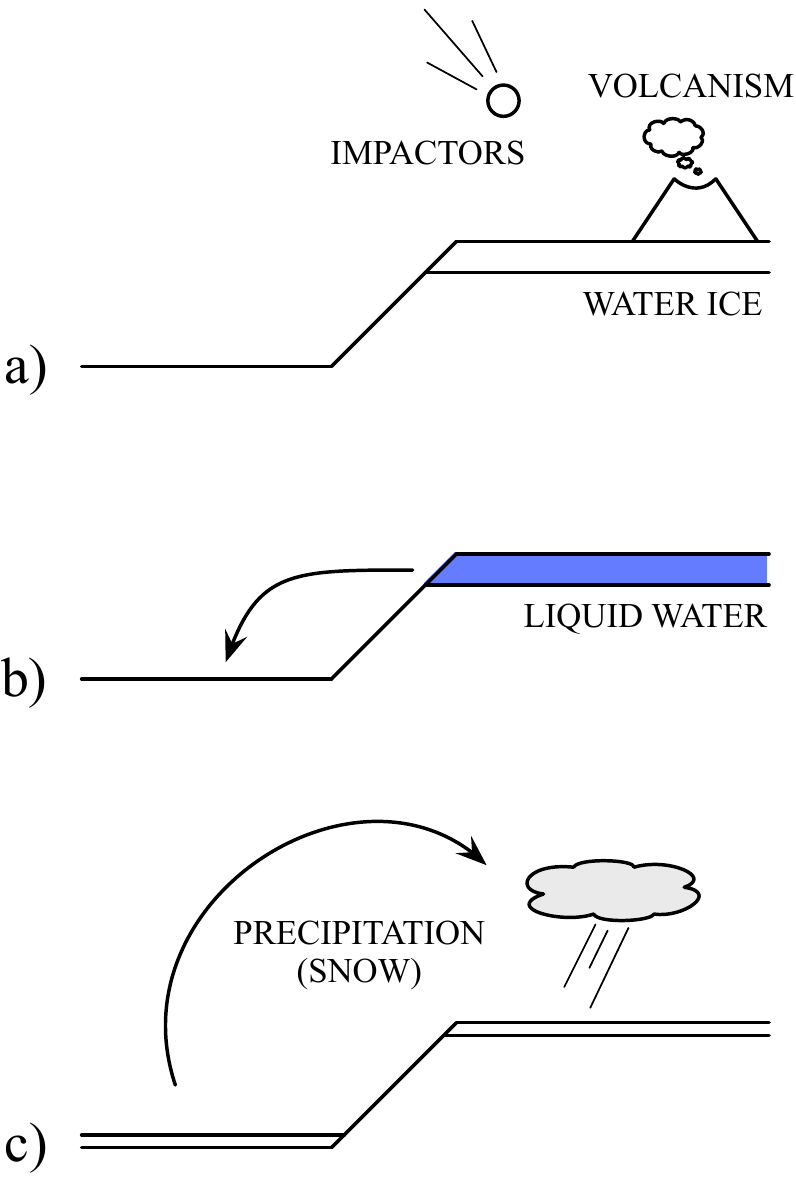}}
	\end{center}
	\caption{Schematic of the effect of periodic melting events under a moderately dense CO$_2$ atmosphere. a) In a steady state, ice deposits are concentrated in the colder highland regions of the planet. b) Impacts or volcanism cause transient ice melting and flow to lower lying regions on short timescales. c) On much longer timescales, ice is once again transported to highland regions via sublimation and light snowfall.}
	\label{fig:impact}
\end{figure}

In our simulations, under the denser CO$_2$ atmosphere that would be expected during Tharsis rise formation, ice was deposited and stabilized in the Noachian highland regions due to the adiabatic cooling effect. Under these circumstances, heating due to impacts could cause extensive melting and hence erosion in exactly the regions where the majority of valley networks are observed \citep{Fassett2008a,Hynek2010}. Transient melting would transport water to lower lying regions, but once temperatures again dropped below the freezing point of water, the slow transport of ice to the highlands via sublimation and snowfall would recommence. As in our simulations that started with ice at low altitudes, the climate system would then return to a state of equilibrium on much longer timescales. Figure \ref{fig:impact} shows a schematic of this process.

Although our results provide a possible solution to one of the problems associated with impact-dominated erosion, the impact hypothesis remains controversial in light of some other observations. For example, young, very large impact craters, such as the Amazonian-era Lyot, have few visible effects of regional or global-scale melting \cite{Russell2002}. In addition, it is still unclear whether the long-term precipitation rates in the highlands predicted by our simulations would provide sufficient H$_2$O deposition to cause the necessary valley network erosion. Further study of hydrology, climate and erosion rates under the extreme conditions expected post-impact are therefore needed to assess the plausibility of this scenario in detail.

\subsection{Volcanism}
\label{subsec:volcanism}

Another potential driver of climate in the Noachian that we have neglected in these simulations is volcanism. As well as causing substantial emissions of CO$_2$, late-Noachian volcanic activity may have influenced the climate via emission of sulphur gases (SO$_2$ and H$_2$S) and pyroclasts (dust/ash particles) \citep{Halevy2007,Halevy2012,Kerber2012}. One-dimensional radiative-convective studies have estimated the sulphur gases to have a potential warming effect of up to 27-70 K \citep{Postawko1986,Tian2010}, while dust is most likely to cause a small (2-10 K) amount of warming, depending on the particle size and vertical distributions \citep{Forget2012}.

As mentioned in the introduction, \cite{Tian2010} argued that the net effect of volcanism should be to cool the early climate, due to the rapid aerosol formation and hence anti-greenhouse effect that could occur after sulphur gas emission. However, their conclusions were based on the results of one-dimensional climate simulations performed without the effects of CO$_2$ clouds included. As described in \cite{Forget2012}, carbon dioxide clouds have a major impact on the atmospheric radiative budget by raising planetary albedo and increasing downward IR scattering. As sulphur aerosols would generally be expected to form lower in the atmosphere, it is therefore likely that their radiative effects would be substantially different in a model where clouds were also included. In particular, if aerosols formed in regions already covered by thick CO$_2$ clouds, they would likely cause a much smaller increase in planetary albedo than if high clouds were absent. Furthermore, sulphate aerosol particles in the CO$_2$-condensing region of an early Martian atmosphere would be potential condensation nuclei for the formation of the CO$_2$ ice cloud crystals, which would further influence the distribution of both aerosol and cloud particles. An extension of the work presented here that included the chemistry and radiative effects of sulphur compounds would therefore be an extremely interesting avenue of future research.

\subsection{Basal melting}
\label{subsec:basal}

It has been proposed that at least some Noachian fluvial features could have formed as a result of the basal melting of glaciers or thick snow deposits \citep{Carr1983,Carr2003}. Geological evidence suggests that throughout much of the Amazonian, the mean annual surface temperatures of Mars have been so cold that basal melting does not occur in ice sheets or glaciers \cite[e.g., ][]{Fastook2008,Head2010,Fastook2011}. However, the documented evidence for extensive and well-developed eskers in the Dorsa Argentea Formation indicates that basal melting and wet-based glaciation occurred at the South Pole near the Noachian-Hesperian boundary. Recent glacial accumulation and ice-flow modeling \citep{Fastook2012} has shown that to produce significant basal melting for typical Noachian-Hesperian geothermal heat fluxes (45-65 mW m$^{-2}$), mean annual south polar atmospheric temperatures of -50 to -75 $^\circ$C are required (approximately the same range as we find in our 1~bar simulations). If geologically based approaches to constraining south polar Noachian/Hesperian temperatures such as this prove to be robust, this provides further evidence in favour of a cold and relatively dry early Martian climate.

\subsection{Conclusions}

Our results have shown that early Mars is unlikely to have been warm and wet if its atmosphere was composed of CO$_2$ and H$_2$O only, even when the effects of CO$_2$ clouds are taken into account. It is possible that other greenhouse gases or aerosols due to e.g. Tharsis volcanic emissions also played a role in the climate. However, given current constraints on the maximum amount of CO$_2$ in the primitive atmosphere (e.g., \cite{Grott2011}; see also \cite{Forget2012} for a detailed discussion), it seems improbable that there was any combination of effects powerful enough to produce a steady-state warm, wet Mars at any time from the mid-Noachian onwards. Despite this, in climates where global mean temperatures are below zero there are still effects that can contribute to transient melting and hence erosion. Simulating the water cycle in 3D has shown that even when the planet is in a state of thermal equilibrium, with mean temperatures as low as $\sim$230~K, seasonal and diurnal warming can cause some localised melting of H$_2$O, although the amount that can be achieved depends strongly on the assumed surface albedo and thermal inertia.

As the era of Tharsis formation / volcanism (and hence of increased CO$_2$ atmospheric pressure) that we have modelled here occurred at a time of elevated meteorite bombardment, impacts will also have repeatedly caused melting of stable ice deposits in the Noachian highlands. Even if post-impact rainfall was relatively short-term, the fact that ice always returned to higher terrain due to the adiabatic cooling effect means that each impact could have created significant amounts of meltwater in the valley network regions. Volcanic activity, particularly that associated with formation of the Tharsis bulge and Hesperian ridged plains \citep{Head2002}, could also have caused discrete warming events in theory, although it remains to be demonstrated how the anti-greenhouse effect due to sulphate aerosol formation could be overcome in this scenario.

Although we believe these simulations have probably captured the main global features of the steady-state Martian climate in the late Noachian, there are clearly a number of other interesting effects that could be investigated in future. We have not included dust or volcanic emissions in our simulations. Both of these are likely to have had an important effect during the Noachian, and adding them in future would allow a more complete assessment of the nature of the early water cycle. Modelling the effects of impacts directly in 3D could also lead to interesting insights, although robust parameterization of many physical processes across wide temperature ranges and short timescales would be necessary to do this accurately. A more immediate application could be to couple the climate model described here with a more detailed subsurface / hydrological scheme. Such an approach could be particularly revealing in localised studies of cases where geothermal processes or residual heating due to impacts may have been important \citep[e.g., ][]{Abramov2005}. In future, integration of 3D climate models with specific representations of warming processes (both local and global) should allow a detailed assessment of whether the observed Noachian fluvial features can be reconciled with the cold, mainly frozen planet simulated here.

\ack This article has benefited from discussions with many researchers, including Jim Kasting, Nicolas Mangold, William Ingram, Alan Howard, Bob Haberle and Franck Selsis.

\bibliographystyle{plainnat}

\label{lastpage}

\end{document}